\documentclass[12pt, singlespacing,nohyperref]{article}
\usepackage[utf8]{inputenc}
\usepackage{authblk}
\usepackage{mathtools}
\usepackage{array}
\usepackage{caption}
\usepackage{geometry}
\usepackage{graphicx}  
\usepackage[colorlinks=true,citecolor=red,linkcolor =blue]{hyperref}
\usepackage[utf8]{inputenc} 
\usepackage[12pt]{moresize}
\usepackage[T1]{fontenc} 
\usepackage[stable]{footmisc}
\usepackage{changepage}
\usepackage[acronym,nomain]{glossaries}              
\usepackage{tcolorbox}
\tcbuselibrary{theorems}
\newtcbtheorem[number within=chapter]{mytheo}{Subtle Detour}%
{colback=blue!5,colframe=blue!35!black,fonttitle=\bfseries}{th}

\usepackage[backend=bibtex, style=ieee,    dashed=false]{biblatex} 

\usepackage[autostyle=true]{csquotes} 
\usepackage{amssymb}
\usepackage{amsmath}

\RequirePackage{geometry}
\title{Exact Low-Energy Solution for Critical Fermi Surfaces}
\author{Tomer Ravid\footnote{Corresponding author: tomeravid@gmail.com} }
\author{Tom Banks}
\affil{NHETC, Rutgers University, Piscataway, NJ}
\date{August 2022}

\begin{document}
\maketitle

\begin{abstract}
We derive multidimensional bosonization directly from the electron gas in a low-energy, low momentum regime where $\omega\gg \frac{k^2}{k_F}$, such that the dispersion can be linearized. To reach this limit, the Fermi momentum and the number of patches are scaled simultaneously keeping the width of each patch finite. We apply this to obtain an exact low-energy solution of the problem of a Fermi surface coupled to a gapless boson, free of disorder and electron-electron scattering. Contrary to claims in the literature, we show that the bosonized theory exactly reproduces the $\omega^{2/3}$ of electrons, previously obtained in large-$N$ theories. We argue that correction to the self-energy due to tangential dispersion are subdominant at sufficiently low energies such that $v_F k\gg \left(\frac{g^4 v_F}{k_F}\right)^{1/3} \omega^{2/3}$, where $g$ is the coupling constant.
\end{abstract}

\newpage

\tableofcontents
\newpage

\section{\label{sec:level1}Introduction}
A Fermi surface is a highly singular object. Electrons at each patch of the Fermi surface lead to a separate resonance, resulting in a continuum of poles in the Green's function. Typically, Fermi surfaces are described by the Landau Fermi liquid theory, which treats the phase-space density $n_{\boldsymbol{k}}(\boldsymbol{x},t)$, and thus the Fermi surface, as a semiclassical variable satisfying a classical kinetic equation and exhibiting no fluctuations. However, when a Fermi surface is coupled to another highly singular object\textemdash namely, an order parameter in the vicinity of a quantum phase transition\textemdash fluctuations in the order parameter turn into fluctuations in the Fermi surface, and deviations from Fermi liquid behavior are observed (see \cite{experiment} and references). This is similar to way in which mean theory becomes invalid in the vicinity of a \textit{classical} phase transition\textemdash the transition from Landau theory to Landau-Ginzburg theory is thus analogous to the transition from Landau Fermi liquid theory to a non-Fermi liquid theory.

The important property of the critical order parameter is that it is a massless boson. There is another mechanism that can generate massless bosons in real situations: a gauge field is naturally massless. However, for conreteness we will focus on the case of an order parameter. We will consider two spatial dimensions as this leads to the most dramatic deviations from Fermi liquid behavior. The simplest model of a critical Fermi surface consists of noninteracting free fermions $\psi$ coupled to an order parameter $\phi$, viz.
\begin{align}\label{mic}
    S\left[\psi,\psi^{\ast},\phi\right]=&\int dt \int d^2p \text{ } \psi^{\ast}(\boldsymbol{p},t)\left[i\partial_t-\varepsilon(\boldsymbol{p})\right]\psi(\boldsymbol{p},t) \nonumber\\
    &\nonumber+g\int dt\text{ }d^2 x\text{ } \phi(\boldsymbol{x},t)\left|\psi(\boldsymbol{x},t)\right|^2\\
    &+\int dt\text{ } d^2 x \left[\dfrac{1}{2}\left(\partial_\mu \phi \right)^2-\dfrac{1}{2} m^2_0 \phi^2\right]
\end{align}
Here, the bare boson mass $m_0$ is needed in order to tune the total mass to zero against renormalization by the interaction with the fermions. For simplicity, here and throughout, we are ignoring the spin degrees of freedom of the electrons, as those will have no effect on the results derived and on the arguments leading to them. We are using units in which the speed of the boson (which clearly differs from the speed of light in spite of the relativistic notation) is unity to simplify the notation. 
The action \eqref{mic} appears to be rather simple. It ignores the effects of disorder and electron-electron interactions. We make no claim that such effects are not important in experiment, but merely treat \eqref{mic} as a minimal model for critical Fermi surfaces.

In spite of the apparent simplicity of the action \eqref{mic}, it has proven difficult to tackle using standard techniques. It becomes strongly coupled at low energies\cite{polchinski}. Thus, perturbation theory is inapplicable. Therefore, a large-$N$ theory for crticial-Fermi surfaces was introduced in \cite{polchinski}, and the leading order term in the $1/N$ expansion was computed. The results are interesting: The boson becomes overdamped and rapidly decays at low energies, at a rate scaling as $\frac{\left|\omega\right|}{\left|\boldsymbol{k}\right|}$. The elecrons are also unstable: There remains a pole at $\omega=0$, i.e., a Fermi surface, but the poles at nonzero energy are destroyed by the interaction with the gapless boson, and decay at a rate proportional to $\omega^{2/3}$. The decay of electrons agrees with experiment, at least qualitatively\cite{experiment}. However, in \cite{sung-sik-lee} it is shown that higher order terms in  the large-$N$ expansion \textit{diverge} at low energies, so they are clearly not negligible!

In \cite{Sachdev}, a model is proposed in an attempt to improve the behavior of the large-$N$ expansion. The model assumes that, in addition to the large number of electron species in the standard large-$N$ theory, there is a large number of boson species\textemdash and that the coupling between each boson and each pair of electrons is a random number obeying some distribution. This model has a strongly self-averaging behavior, and the $\frac{1}{N}$ corrections average to zero, leaving one with the same results obtained by considering single-loop diagrams in the standard theory. However, it remains unclear why the additional structure assumed by this model should not change the physics of the problem, and why such assumptions should yield no important deviation from reality. 

It is thus evident that a fundamental understanding of critical Fermi surfaces requires a more direct approach, and such an approach we will attempt to provide. 

In this paper, we study the problem assuming nothing but ordinary electrons and ordinary critical bosons, and without any kind of an approximation other than a particular low energy, low momentum limit. We argue for what might appear ``too good to be true''\textemdash that, at least in the absence of disorder and electron-electron interactions, critical Fermi surfaces are not merely less complicated than previously thought, but they are in fact simple enough to be solved exactly. We argue that the only contribution to the solution is due to simple fermionic loops with no internal decorations\textemdash the same diagrams that contribute in the large-$N$ theory, in spite of the absence of a large-$N$ approximation. This vast difference from previous studies owes to a nontrivial way of taking the low-energy limit of the Fermi surface. The nontrivial low energy limit guarantees the irrelevance of the quadratic term in the electron's dispersion, which is the main source of difficulty faced by previous studies.

\section{Linearizing the Fermions Near the Fermi Surface}\label{linearization}
The Fermi surface is a surface $k_F(\chi)$ satisfying
\begin{align}
    \varepsilon\left(k_F(\chi) {\hat{\boldsymbol{x}}}_{\perp} \right)=0
\end{align}
where $\boldsymbol{\hat{x}}_{\perp}=\left(\cos\chi,\sin \chi\right)$ is the normal to the Fermi surface, and $\chi$ is an angle labelling a patch of the Fermi surface through the direction of its normal (for a spherical Fermi surface this coincides with the polar angle). If one is interested in low-energy degrees of freedom, then one should consider electrons in the vicinity of the Fermi surface. Naively, this can be done by expanding $\varepsilon(\boldsymbol{p})$ to linear order near $\left|\boldsymbol{p}\right|=k_F(\chi)$. This is the approach taken, for example, in \cite{shankar}. However, this leads to a dispersion
\begin{align}\label{naive-linearization}
    \varepsilon(\boldsymbol{p})\approxeq v_F (\chi) \left(\left|\boldsymbol{p}\right|-k_F (\chi)\right).
\end{align}
giving rise to a singular action in momentum space, and thus to an apparently nonlocal action in position space. The singular behavior arises because the dispersion has been linearized around a \textit{surface} rather than around a \textit{point}. Linear dispersion around any \textit{specific} point entails no such nonlocality. The appropriate way to obtain a low-energy local theory of Fermi surface is thus to divide the Fermi surface into patches, and then linearize around the Fermi point separately for each patch. 

The linearization program will result in an infinite collection of massless fermions linearly dispersing in the direction normal to the Fermi surface. A similar approach is presented and criticized e.g. by Sachdev in \cite{sachdev-text}. However, Sachdev's discussion assumes a continuum limit in which the width of each patch is zero. As a result, each patch becomes a strictly one-dimensional line, and (the physically important) scattering tangent to the Fermi surface is thereby ignored, as is the curvature of the Fermi surface. The conventional wisdom\cite{polchinski}\cite{sung-sik-lee}\cite{Sachdev}\cite{sachdev-text} is thus that one must expand the dispersion to quadratic order, so that the dispersion depends on the tangential momentum, and each patch becomes a full and rather complicated two-dimensional field theory. This has been implemented within a multi-patch framework similar to our own in \cite{goldman}. Here, we argue that the loss of tangential scattering in the naively linearized model is not due to the linearization of the dispersion, but instead due to the continuum limit of the patches. Instead, as one might naively expect, the quadratic dispersion is irrelevant at low momenta compared to the linear dispersion, but only so long as one takes a non-naive continuum in which each patch maintains a finite width. The result is that each patch is a ``one-and-a-half-dimensional field theory'': it obeys a one-dimensional dispersion, but its momentum lies on a two-dimensional plane. This is a crucial point: one-dimensional massless fermions are exactly soluble (e.g. as shown by Schwinger in the context of QED\cite{schwinger}), and it will be shown that the the ``one-and-a-half-dimensional'' patch inherits this exact solubility. 

To linearize near the Fermi momentum, note that a general momentum can be written as a sum of the Fermi momentum of a patch and a deviation from that Fermi momentum
\begin{align}
    \boldsymbol{p}=k_F (\chi) \boldsymbol{\hat{x}}_{\perp}(\chi)+\boldsymbol{k}.
\end{align}
One can thus linearize the dispersion around the point by taking the deviation $\boldsymbol{k}$ to be small
\begin{align}
     \varepsilon(\boldsymbol{k},\chi)&\approxeq\left.\dfrac{\partial \varepsilon}{\partial \boldsymbol{k}}\right|_{\boldsymbol{k}=0,\chi}\cdot{\boldsymbol{k}} \nonumber \\
     &=v_F (\chi) \boldsymbol{k}\cdot\boldsymbol{{\hat{x}}}_{\perp}(\chi).
\end{align}
Note that even though the resulting dispersion is only in the normal direction, there is no underlying assumption that $\boldsymbol{k}$ is normal to the Fermi surface, but merely that it is small in magnitude compared to the Fermi momentum. The momentum is transferred to the fermions by the critical boson\textemdash and provided that that the momentum carried by the boson is sufficiently small compared to the Fermi momentum, the linear approximation holds, and the patch behaves as a ``one-and-a-half-dimensional'' fermion with two dimensional momentum but one-dimensional dispersion\footnote{For further assessment of the validity of the lineraization of the dispersion, see Section \ref{discussion}}. This dispersion is highly singular and will be shown to lead to some of the singular results obtained in large-$N$ theories of critical Fermi surfaces.

Since there is nothing special about the patch, one can split the entire Fermi surface into a set of $N$ patches $\chi_n=\frac{2\pi n}{N}$, and write every momentum as the sum of the Fermi momentum of the closest patch and its deviation from that patch $k\lesssim\frac{\pi k_F}{N}$. The fermion action in \eqref{mic} can thus be linearized as
\begin{align}\label{after}
    S_f\approxeq &\sum^{N}_{n=0}\int dt \int_{k\leq\frac{\pi k_F}{N}} d^2k \text { }\psi^{\ast}\left(\boldsymbol{k},\chi_n,t\right)\left[i\partial_t-v_F \left(\chi_n\right) k \cos \left(\chi_n-\theta\right) \right]\psi\left(\boldsymbol{k},\chi_n,t\right),
\end{align}
with $\theta$ the polar angle of $\boldsymbol{k}$. Unlike \eqref{naive-linearization}, the action \eqref{after} is manifestly local, because the normal momentum $k_{\perp}\equiv k\cos(\theta-\chi)$ Fourier transforms to the normal derivative $\partial_{\perp}\equiv\cos \chi \partial_x +\sin \chi \partial_y$.

\begin{figure}[t]
    \centering
    \captionsetup{justification=centerlast}
    \includegraphics[width=16cm]{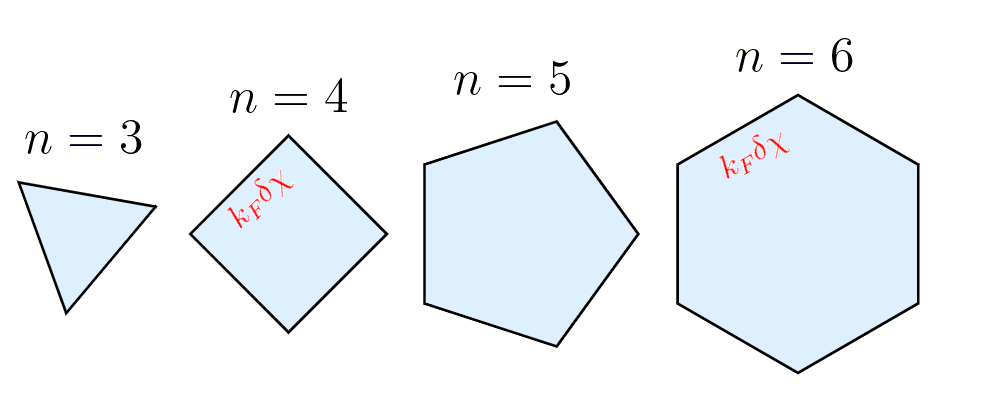}
    \caption{In the double scaling limit, one approximates the Fermi surface as a polygon with each side having a fixed width $k_F \delta \chi$, and takes the limit in which the number of sides approaches infinity.}
    \label{fig:polygon}
\end{figure}

The next natural step is to take the continuum limit of \eqref{after}. However, this limit must be taken with some care: if the width of a patch were strictly zero (as in \cite{sachdev-text}), then arbitrarily small momentum transfer would make an electron hop away from its patch, and by ignoring such inter-patch hopping one loses track of all scattering tangent to the Fermi surface. Instead, the width of a patch should be viewed as an upper cutoff on the tangential momentum (it is the momentum an electron must acquire to leave a patch). One must take a double scaling limit in which the the Fermi momentum is taken to infinity simultaneously with the number of patches, so that the cutoff $\Lambda=k_F \delta \chi$ is fixed. In other words, 
\begin{align}\label{double-scaling}
    \Lambda=\dfrac{2\pi k_F}{N}=\text{const}.
\end{align} 
This limit can be visualized as follows. One approximates the Fermi surface as a polygon with each side having a fixed width $k_F \delta\chi$, and takes the limit in which number of sides approaches infinity. This is depicted for a Fermi sphere in figure \ref{fig:polygon}.

In this limit, $\chi$ becomes a continuous variable. After rescaling $\psi$ in such a way that $\left|\psi\right|^2$ represents charge \textit{density} per patch rather than charge, the action takes the form
\begin{align}\label{ferm-lin}
    S=\int^{2\pi}_0 d\chi \int dt \text{ } d^2 x \psi^{\ast} (\boldsymbol{x},t,\chi)\left[i\partial_t-i v_F(\chi)\partial_{\perp}-g\phi(\boldsymbol{x},t) \right]\psi(\boldsymbol{x},t,\chi)+S_b\left[\phi\right],
\end{align}
with $S_b[\phi]$ the free action of the critical boson. The free ($g=0$) field equation derived from \eqref{ferm-lin} simply displaces the fermion field at each patch in the outgoing (thus positive) normal direction. Assuming inversion symmetry, $v_F(\chi)=v_F(\chi+\pi)$. By analogy with the one-dimensinoal electron gas, one can combine the incoming and outgoing modes from antipodal patches into a single Dirac spinor, defining:
\begin{align}
    \Psi \left(t,\boldsymbol{x},\chi\right)\equiv
    \begin{pmatrix}
    &\psi\left(t,\boldsymbol{x},\chi\right)\\
    &\psi \left(t,\boldsymbol{x},\chi+\pi\right)
    \end{pmatrix},
\end{align}
so that the action can be rewritten as 
\begin{align}\label{weyl-lin}
    S&=\int^{\pi}_0 d\chi \int dt\text{ } d^2 x \text{ } \Psi^{\dagger}\left(t,\boldsymbol{x},\chi\right)\left[i\partial_t-i v_F \left(\chi\right) \sigma_{z} \partial_{\perp} + g\phi(\boldsymbol{x},t) \right]\Psi\left(t,\boldsymbol{x},\chi\right)+S_b\left[\phi\right] \nonumber\\
    &=\int^{\pi}_0 d\chi \int dt \text{ } d^2 x \text{ } \bar{\Psi}\left(t,\boldsymbol{x},\chi\right) \left[i \gamma^{a} \partial_{a}+g\gamma^{0} \phi(\boldsymbol{x},t)\right]\Psi\left(t,\boldsymbol{x},\chi\right)+S_b\left[\phi\right],
\end{align}
where we have introduced shorthand qausi-relativistic notation: $\partial_{a}\equiv\left(\partial_t,v_F(\chi)\partial_{\perp}\right)$, $\gamma^{a}=\left(\sigma_x,\sigma_y\right)$. The Minkowski metric raising and lowering indices is just $\eta_{ab}=\text{diag}\left(1,-1\right)$, so the entire dependence on the Fermi velocity (which, contrary to case of the speed of light, is physically important and nontrivial) is contained in the derivative vector. It has thus been established that the Fermi surface is equivalent to an infinite collection of linearly dispersing ``one-and-a-half-dimensional'' Weyl fermions, to which the critical boson couples as an electric potential. Note that in spite of the normal dispersion, contrary to what is argued about the naive continuum limit of \cite{sachdev-text}, the curvature of the Fermi surface is not ignored. Rather, it is directly embodied in the fact that normals of different patches are not parallel\textemdash thus the electrons of different patches move in different directions. We will show that this is sufficient in order to capture the physics of effects normally attributed to the curvature the Fermi surface: the standard electron and boson Green's functions.

\section{Exact Low-Energy Solubility of the Linearized Action}\label{direct}
It is well known that the problem of one-dimensional Weyl fermions coupled to an external field can be solved exactly, as originally shown by Schwinger \cite{schwinger}. Since each patch of the Fermi-surface behaves at low momenta like a ``one-and-a-half-dimensional Weyl fermion,'' it is to be expected that the low-momentum theory can be solved exactly on similar grounds. Indeed, direct path integration establishes that this is the case. It further establishes that if the dispersion is linear, then the only contribution to the solution comes from simple fermionic loops\textemdash explaining why large-$N$ models give seemingly reasonable answers.

The linearized action \eqref{weyl-lin} is quadratic in fermion fields and so the fermion functional integral can be computed exactly.  According to the linearized action \eqref{weyl-lin}, the expression for multi-fermion correlators can be written in terms of the Green's function of the fermion in an external field
\begin{align}\label{green}
\left[i\gamma^{0}\gamma^{a}\partial_{a} + g \phi (t,\boldsymbol{x})\right] G\left(t,t^{\prime}, \boldsymbol{x}, \boldsymbol{x^{\prime}} , \chi, \chi^{\prime}\right) = \delta \left(t - t^{\prime}\right) \delta^2 \left(\boldsymbol{x} - \boldsymbol{x^{\prime}}\right) \delta (\chi - \chi^{\prime}), 
\end{align} 
functionally averaged over the effective action
\begin{align} 
S_{\text{eff}} \equiv S_b [\phi] + {\rm Tr \ln}\left[i\gamma^{0}\gamma^{a}\partial_{a} + g \phi (t,\boldsymbol{x})\right],
\end{align} 
where the same quasi-relativistic notation from \eqref{weyl-lin} has been used.

The determinant can be computed from the Green's function via the formula
\begin{align} \dfrac{\delta {\rm Tr\ ln}\ \left[i\gamma^{0}\gamma^{a}\partial_{a} + g \phi (t,\boldsymbol{x})\right]}{
\delta \phi (t, \boldsymbol{x})} =  g {\rm Tr}\ G(x,x) . \end{align}
The key to the solution of these models is the same as the one Schwinger\cite{schwinger} used to solve massless quantum electrodynamics in $1 + 1$ dimensions. Equation \eqref{green} states that the external field Green's function moves right or left while acquiring a phase depending on $\phi$. It thus has an explicit analytic solution of the form of a free Green's fnuction $G_0$ with an ``Aharonov-Bohm phase'' $B$
\begin{align}\label{AB-phase} G\left(t,t^{\prime}, \boldsymbol{x}, \boldsymbol{x^{\prime}} , \chi, \chi^{\prime}\right) = G_0 (t,t^{\prime}, \boldsymbol{x}, \boldsymbol{x^{\prime}}) e^{- i \int_0^{\pi} d\chi\ B\left(t,t^{\prime}, \boldsymbol{x}, \boldsymbol{x^{\prime}}, \chi\right) } \delta (\chi - \chi^{\prime}) . \end{align}  Here, $G_0$ is the $\chi$-independent $g = 0$ solution and
\begin{align} \gamma^0\gamma^{a}\partial_a B\left(t,t^{\prime}, \boldsymbol{x}, \boldsymbol{x^{\prime}} , \chi\right) = -g \phi (t, \boldsymbol{x}) . \end{align}  In addition, we must have
\begin{align} \int_0^{\chi} B(t,t, \boldsymbol{x}, \boldsymbol{x} , \chi) = 0 .\end{align} 

The last condition implies that, {\it in the expression for the logarithm of the determinant only the term linear in the first derivative of $B$ at coinciding points survives.}  The Taylor series truncates because $G_0$ has only a pole at coinciding points.  As a consequence, the logarithm of the determinant is quadratic in $\phi$, which means that it is given exactly by the one loop fermion vacuum polarization bubble!   This is not a large-$N$ expansion, but an exact property of this class of linearly dispersing critical Fermi surfaces.  The expressions for fermion propagators are all linear functionals of $\phi$, which means that they can be evaluated exactly because the $\phi$ action is quadratic.

Every single calculation of this paper can be performed directly from the path integral of the resulting effective action, which is just a Hertz-Millis action\cite{hertz}\cite{millis}. However, in what follows we will rewrite the theory in terms of bosonized variables that are not strictly necessary computationally, but are nevertheless invaluable conceptually. 

\section{Exactness of Bosonization}\label{section3}
Eq. \eqref{AB-phase} states that the sole effect of the order parameter $\phi$ on the electrons is to shift their phases linearly. Therefore, it is intuitively clear that if one were to perform some kind of a ``polar decomposition'' of the electron field into chiral phases and current densities, the current densities should decouple from the order parameter while the phases would behave like an infinite collection of bosonic fields (one per patch) linearly coupled to the order parameter. Such a description would be a natural two-dimensional generalization of the bosonized description of the one-dimensional Luttinger liquid, in which the basic variable is the bosonic phase field associated with electron-hole excitations. 

\begin{figure}
    \centering
        \captionsetup{justification=centerlast}
    \includegraphics[width=15cm]{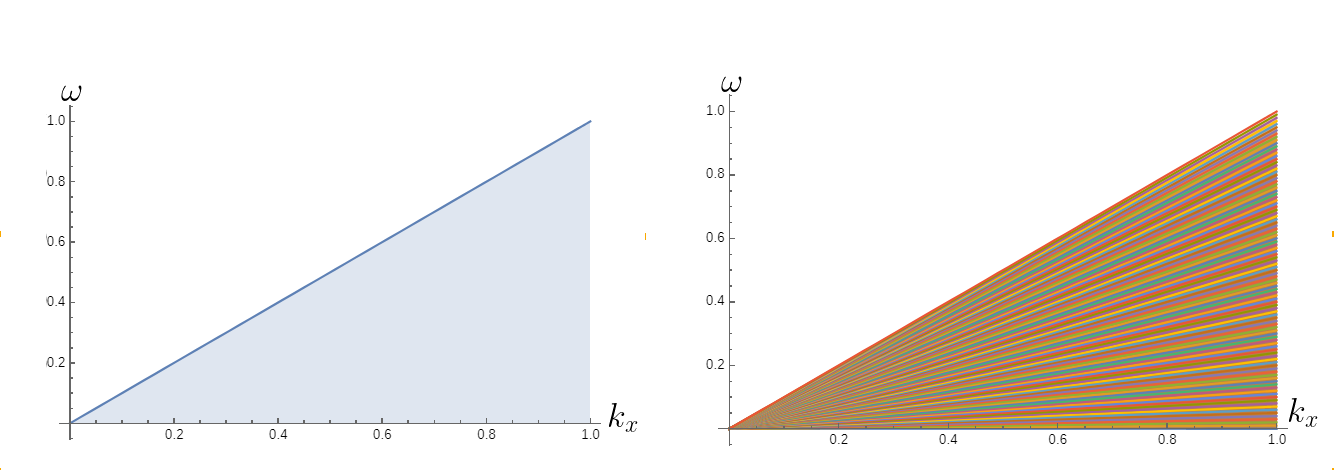}
    \caption{Haldane's idea is essentially to fill the electron-hole continuum (left) with an infinite collection of linearly dispersing excitations of different slopes (right), corresponding to electron-hole pairs at different patches.}
    \label{fig:2d-bosonization}
\end{figure}

Indeed, such a higher-dimension bosonized formalism was proposed by Haldane\cite{haldane} as a general framework for fluctuating Fermi surfaces. Haldane proposed that that a Fermi surface is equivalent to an infinite collection of linearly dispersing bosons, each corresponding to electron-hole pairs created at a different patch\textemdash essentially filling the electron-hole continuum with an infinite number of lines (see figure \ref{fig:2d-bosonization}).

Lawler at al. applied Haldane's formalism to the problem of critical Fermi surfaces\cite{lawler}, and showed that it reproduces the results from the random phase approximation. In spite of these impressive results, the exactness of linear bosonization at low energies has remained far from accepted in the literature. Even proponents of the  bosonization approach to non-Fermi liquids often assume that the quadratic term in the electron's dispersion is relevant and leads to nonlinear corrections to the bosonized action, and/or that the quadratic bosonzed action is merely a reformulation of the random phase approximation\cite{sohn}\cite{goldman}. By contrast, the result of Section \ref{linearization} is that there is a limit in which the electronic dispersion is fully linear, and in which each patch behaves similarly to a one-dimensional Weyl fermion. Therefore, it should not be surprising that in this limit, one can bosonize the Weyl fermion of each patch similarly to the one-dimensional case, leading to an exact, microscopic derivation of the bosonized model.  

As previously stated, the intutive idea is to decompose the eletcron field into a phase and a magntiude. This idea is essentially correct, and would lead to the desired result; yet it is hard to make it rigorous, since the electron field is a Grassmann coordinate, and because of the chiral anomaly. Instead of decomposing the electron field into a magntiude and a phase (i.e., shifting the phases to zero), consider an \textit{arbitrary} local shift of the phase. This amounts to defining a new field $\Gamma$ as

\begin{align}\label{trans}
    \Gamma\left(\boldsymbol{x},t,\chi\right)\equiv\exp\left[-i\sigma_z \overline{\zeta}\left(\boldsymbol{x},t,\chi\right)-i\overline{\eta}\left(\boldsymbol{x},t,\chi\right) \right]\Psi \left(\boldsymbol{x},t,\chi\right).
\end{align}
for some phase functions $\overline{\zeta}$ and $\overline{\eta}$. Plugging this into the classical action gives

\begin{align}\label{weyl-lin}
    S&\rightarrow\int^{\pi}_0 d\chi \int dt \text{ } d^2 x \text{ } \bar{\Gamma}\left(t,\boldsymbol{x},\chi\right) \left[i \gamma^{a} \left(\partial_{a}+i\epsilon_{ab}\partial^{b}\overline{\zeta}+i\partial_{a}\overline{\eta}\right)+ g\gamma^{0} \phi(\boldsymbol{x},t)\right]\Gamma\left(t,\boldsymbol{x},\chi\right)+S_b\left[\phi\right],
\end{align}
 Since the entire effect of the critical field on the electron field is encoded in the phase of \eqref{AB-phase}, it is natural to suspect that the critical boson can be decoupled from the electrons by appropriately choosing the phase shifts. This is indeed the case, because by the Helmholtz decomposition \textit{any} two-dimensional vector field can be written as the sum of ``irrotational flow'' $\partial_{a}\overline{\eta}$ and ``incompressible flow'' $\epsilon_{ab}\partial^{b}\overline{\zeta}$. Hence, by choosing $\overline{\eta}$ and $\overline{\zeta}$ such that
 \begin{align}\label{cancellation}
     \epsilon_{ab}\partial^{b}\overline{\zeta}\left(\boldsymbol{x},t,\chi\right)+\partial_{a}\overline{\eta}\left(\boldsymbol{x},t,\chi\right)=-g\phi\left(\boldsymbol{x},t\right) \delta^{0}_a,
 \end{align}
the interaction term disappears. It thus appears that by integrating the fermions out, one recovers the free boson action, with no effect from the fermions! However, there is an additional effect of the fermions for which the classical action does not account: by the chiral anomaly, the measure of the path integral changes under the transformation \eqref{trans}, and the transformation of the measure will give rise to an additional term in the action. Computing the transformation of the measure is similar to the standard Fujikawa method for 1+1D fermions in an external electromagnetic potential. The main difference is that the effective potentials, which depend on $\phi$ and on $\overline{\zeta}$, change under the chiral transformation. Hence, one should break the transformation into infinitesimal blocks and compute the change of the measure under each one of them. This is done in the one dimensional case in \cite{chir}, where the full details of the derivation can be found. Here, we will just outline the steps.

Let $\alpha$ be a variable parameterizing the ``stage'' of the transformation. Similarly to \cite{chir}, we will choose $\alpha$ such that $\alpha=0$ labels the initial stage, which is just the original action \eqref{weyl-lin}, in which the effective potential is  $g\phi\delta^{a}_0$, and $\alpha=0$ labels the final state, in which the fermions have been fully decoupled and thus the effective potential vanishes. The effective potential at any intermediate state is
\begin{equation}
    A^{a}(\alpha)=\left(1-\alpha\right) \epsilon^{ab}\partial_{b}\overline{\zeta},
\end{equation}
where \eqref{cancellation} has been used, with the $\overline{\eta}$-term discarded as a gauge transformation which will have no effect on the result. The total measure of the fermions will be a product of the $1+1$-dimensional measures of all patches $\chi$ and all positions in the direction tangent to the Fermi surface $x_{\parallel}\equiv x\sin\chi-y\cos\chi$. One can thus apply the standard 1+1-dimensional Fujikawa formula\cite{fujikawa} to the each patch and each tangential position, and at each infinitesimal stage of the transformation
\begin{align}
    d\overline{\Gamma}\left(x_{\parallel},\chi\right) d\Gamma \left(x_{\parallel},\chi\right)\nonumber=&\exp\left[i\int \dfrac{dt dx_{\perp}}{2\pi v_F(\chi)} \int^{1}_{0} d\alpha  \overline{\zeta}\left(x_{\perp},x_{\parallel},t,\chi\right) \epsilon^{ab} \partial_{a} A_{b} (\alpha)\right]\nonumber\\  &\times d\overline{\Psi}\left(x_{\parallel},\chi\right) d{\Psi}\left(x_{\parallel},\chi\right)\nonumber\\
    = &\nonumber\exp\left[\dfrac{i}{4\pi}  \int \dfrac{dt\text{ } dx_{\perp}}{v_F(\chi)} \tilde{\zeta}\left(t,x_{\perp},x_{\parallel},\chi\right) \partial_{a} \partial^{a} \tilde{\zeta}\left(t,x_{\perp},x_{\parallel},\chi\right)\right]\\
    &\times d\overline{\Psi}\left(x_{\parallel},\chi\right) d{\Psi}\left(x_{\parallel},\chi\right).
\end{align}
The total transformation of the measure $\mathcal{D}\Gamma \mathcal{D}\overline{\Gamma}$ will then be a product over patches and tangential positions. Because of subtleties of the double scaling limit, computing the product requires lattice regularization, so $\chi$ is replaced with some discrete $\chi_n$ and $x_{\parallel}$ with $x_{\parallel,m}$. The resulting transformation is 

\begin{align}\label{anomaly}
    \mathcal{D}\overline{\Gamma}\mathcal{D}\Gamma\nonumber&=\prod_{n,m} d\overline{\Gamma}\left(x_{\parallel,m},\chi_n\right) d\Gamma \left(x_{\parallel_m},\chi_n\right)\\
    &=\mathcal{D} \overline{\Psi} \mathcal{D} \Psi \exp\left[\dfrac{i}{4\pi} \sum_{n,m} \int \dfrac{dt\text{ } dx_{\perp}}{v_F(\chi)} \tilde{\zeta}\left(t,x_{\perp},x_{\parallel,m},\chi_n\right) \partial_{a} \partial^{a} \tilde{\zeta}\left(t,x_{\perp},x_{\parallel,m},\chi_n\right)\right].
\end{align}
The sum looks worryingly divergent in the continuum limit, owing to the absence of an infinitesimal measure. However, observe that in the double scaling limit the Fermi momentum is also ``divergent'' compared to any momentum $k$, and yet it is a parameter of the theory. It turns out that the divergence of the sum is precisely of the same order as the divergence of the Fermi momentum. The double scaling limit necessitates that the discretization of the tangential position is inversely related to the upper momentum cutoff\textemdash which, in turn, is proportional to the width of the patch (i.e., the discretization of $\chi$) via eq. \eqref{double-scaling}:
\begin{equation}
    \delta x_{\parallel} =\dfrac{2\pi}{\Lambda}=\dfrac{2\pi}{k_F \delta \chi},
\end{equation}
in other words, the missing integration measure $\delta \chi \delta x_{\parallel} $ is simply given by the Fermi wavelength:
\begin{equation}
    \delta x_{\parallel} \delta \chi =\dfrac{2\pi}{k_F},
\end{equation}
and plugging this into \eqref{anomaly} gives

\begin{align}
    \mathcal{D}\overline{\Gamma}\mathcal{D}\Gamma\nonumber&=\mathcal{D} \overline{\Psi} \mathcal{D} \Psi \exp\left[i \dfrac{1}{8\pi^2 } \sum_{n,m} k_F\left(\chi_n\right) \delta \chi \delta x_{\perp} \int \dfrac{dt\text{ } dx_{\perp}}{v_F\left(\chi_n\right)} \tilde{\zeta}\left(t,x_{\perp},x_{\parallel,m},\chi_n\right) \partial_{a} \partial^{a} \tilde{\zeta}\left(t,x_{\perp},x_{\parallel,m},\chi_n\right)\right]\\
     &\overset{\text{continuum}}{\rightarrow} \mathcal{D} \overline{\Psi} \mathcal{D} \Psi \exp\left[i \int^{\pi}_0 d\chi \dfrac{k_F(\chi)}{8\pi^2 } \int \dfrac{dt\text{ } d^2x}{v_F(\chi)} \tilde{\zeta}\left(t,\boldsymbol{x},\chi\right) \partial_{a} \partial^{a} \tilde{\zeta}\left(t,\boldsymbol{x},\chi_n\right)\right]
\end{align}
$\tilde{\zeta}$ is a nonlocal functional of $\phi$, defined by \eqref{cancellation}\textemdash or, equivalently, by its curl
\begin{equation}\label{calssical-field-equation}
    \partial_{a}\partial^{a} \overline{\zeta}=-g v_F (\chi)\partial_{\perp} \phi.
\end{equation}
This can be used to simplify the factor of $\partial_{a}\partial^{a} \overline{\zeta}$ in the transformation of the measure, so the resulting path integral is
\begin{equation}
 Z=\int \mathcal{D} \phi \mathcal{D} \overline{\Gamma} \mathcal{D} {\Gamma} \exp\left[i g \int^{\pi}_0 d\chi \dfrac{k_F(\chi)}{8\pi^2 } \int {dt\text{ } d^2x} \tilde{\zeta}\left(t,\boldsymbol{x},\chi\right)\partial_{\perp} \phi \left(t,\boldsymbol{x}\right)+i S_b\left[\phi\right]+i S_{f}\left[\Gamma,\overline{\Gamma}\right] \right],
 \end{equation}
with $S_f$ the free fermion action of $\Gamma$, completely decoupled from $\phi$. The direct dependence on $\overline{\zeta}$ of the term that originated from the anomaly means that this action still depends on $\phi$ nonlocally. Of course, this dependence is  identical to the one in the Hertz-Millis action obtained by the method of Section \ref{direct}. This nonlocality of the effective action can be traced back to the fact that $\overline{\zeta}$ is not an integration variable, but a classical field constrained to solve the classical field equation \eqref{calssical-field-equation}. While the classical field equation is local, its solution depends nonlocally on $\phi$. However, the classical field equation minimizes a quadratic action, and it is well-known that integrating out a field with a quadratic equation (e.g. the electromagnetic field in QED) is equivalent to simply plugging a classical solution into the path integral. The implication is that one can rewrite the nonlocal term as a local term with an additional integration variable $\zeta\left(\boldsymbol{x},t,\chi\right)$:
\begin{align}
     \exp\left[i g \int^{\pi}_0 d\chi \dfrac{k_F(\chi)}{8\pi^2 } \int {dt\text{ } d^2x} \tilde{\zeta}\left(t,\boldsymbol{x},\chi\right) \partial_{\perp} \phi\right]=\int \mathcal{D} \zeta \exp&\left[i\int^{\pi}_0 d\chi \dfrac{k_F(\chi)}{16\pi^2 v_F(\chi) } \int {dt\text{ } d^2x}\left(\partial_{a}\zeta \partial^{a} \zeta\right.\right.\nonumber\\ &\left.+2gv_F(\chi){\zeta}\left(t,\boldsymbol{x},\chi\right)\partial_{\perp} \phi\left(t,\boldsymbol{x}\right)\right) \bigg].
\end{align} 

All in all, integrating out the decoupled fermions $\Gamma$, one finds that a path integral of an infinite collection of bosonic phase fields, coupled to the critical order parameter via a quadratic action
\begin{align}\label{bosonized-action}
    S[\zeta,\phi]=\dfrac{1}{16\pi^2} \int^{\pi}_{0} d\chi \dfrac{k_F(\chi)}{v_F(\chi)} \int dt\text{ } d^2x&\left[\partial_{a} \zeta\left(t,\boldsymbol{x},\chi\right)  \partial^{a} \zeta\left(t,\boldsymbol{x},\chi\right)\right.\nonumber\\
    &\left.+2g v_F(\chi)\phi(t,\boldsymbol{x})\partial_{\perp}\zeta\left(t,\boldsymbol{x},\chi\right)\right]+S_{b}[\phi].
\end{align}
This is nothing but a multi-dimensional bosonized action for the critical Fermi surface\textemdash derived directly and exactly as a low energy limit of the microscopic action \eqref{mic}. 

\section{Exact Boson Green's Function}\label{section4}
The bosonized action \eqref{bosonized-action} is quadratic, and so the correlation functions can be easily computed. Because of the quadraticity of the action, one might assume that it leads to no nontrivial correlators. However, as shown in \cite{lawler}, this action reproduces the non-Fermi liquid Green's functions computed within the large-$N$ approximation. 

The nontriviality of the results derived within an innocuous-looking quadratic action owes to the abundance of poles of the Fermi surface, as alluded in the introduction. This abundance of poles implies, in particular, that an incoming boson will always resonate with some patch of the Fermi-surface, regardless of its specific energy and momentum (but provided that they are within the electron-hole continuum)\textemdash thus, there is always some on-shell state to which the boson can decay. This is clearly an effect of the curvature of the Fermi surface, for if all patches moved in the same direction there would be no multitude of resonances.

The above is the source of Landau damping of the boson, and is seen clearly in the explicit calculation of the boson's Green's function. The basic calculation from the bosonized theory was laid out in \cite{lawler} and is recapitulated in \cite{sohn}, but we would like to stress an interesting result regarding the analytic behavior of the boson self-energy that was not discussed in these references, so we will briefly outline the calculation. 

\begin{figure}
    \centering
        \captionsetup{justification=centerlast}
    \includegraphics[width=10 cm]{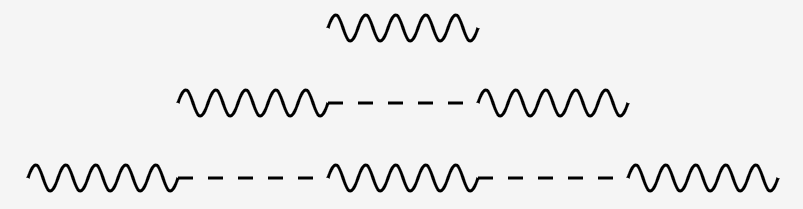}
    \caption{A boson (wavy line) can emit a density wave (dashed line) and re-absorb it any number of times. The amplitude for each such event is some number, and the sum of all such amplitudes is a geometric series, and the amplitude of each emission and reabsorption process if the self-energy.}
    \label{fig:feyn}
\end{figure}

The basic process that contributes to the boson's Green function is the following: the boson can emit an electron-hole pair at each patch, then reabsorb it. This process gives an energy and momentum-depedent number, and can happen any number of times, resulting in a geometric series for the boson's Green's function (see figure \ref{fig:feyn}). The number is just the self energy, which is proportional to the electron-hole field's (i.e., $\zeta$'s) Green's function and to the squared (momentum-dependent) coupling, averaged over all patches:
\begin{align}\label{pi-int-vf}
    \Pi(\omega,\boldsymbol{k})= \dfrac{g^2}{4\pi^2} \int^{\pi}_{0} d\chi \dfrac{k_F(\chi)}{v_F(\chi)} \dfrac{v_F^2 k^2 \cos^2(\theta-\chi)}{\omega^2-v^2_F(\chi) {k^2} \cos^2\left(\chi-\theta\right)+i\epsilon}.
\end{align}
where $\theta$ is the polar angle of momentum, $\boldsymbol{k}=\left(k\cos\theta,k\sin\theta\right)$. This is most easily calculated in the case of a Fermi \textit{sphere}, in which $k_F$ and $v_F$ are independent of $\chi$. In this case, there is a symmetry under simultaneous rotation of $\theta$ and $\chi$, and the result could be computed without loss of generality by choosing coordinates in which $\theta=0$. 

\begin{figure}[t]
    \centering
        \captionsetup{justification=centerlast}
    \includegraphics[width=15cm]{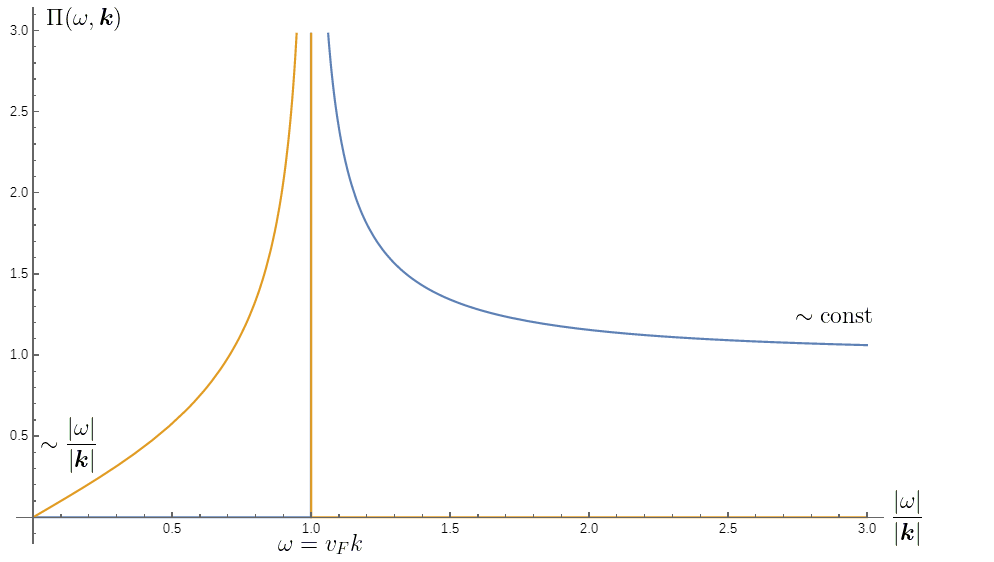}
    \caption{In the case of a Fermi sphere, the boson's self energy starts out linear in $\frac{\left|\omega\right|}{\left|\boldsymbol{k}\right|}$. It has a pole at $\omega=v_F k$, below which it is purely imaginary and above which it is purely real. It then asymptotes to a finite ``mass'' for small momenta/high energies.}
    \label{fig:self-energy}
\end{figure}

When $\omega<v_F k$, the boson's phase velocity is smaller than the Fermi velocity, and there is always some resonant patch satisfying $\omega=\pm v_F k \cos \chi^{\star}$, leading to a pole in the denominator of \eqref{pi-int-vf}. We will call this case the ``subsonic regime.'' The result would be then a sum of the principal value and a contribution from the pole
\begin{align}\label{pi-int2}
    \nonumber \Pi(\omega,\boldsymbol{k})= &\dfrac{k_F g^2}{4\pi^2 v_F} \mathcal{P} \int^{\pi}_{0} d\chi \dfrac{\cos^2\chi}{\dfrac{{\omega}^2}{{v_F^2 k}^2}- \cos^2\chi}\\&+i \dfrac{k_F g^2}{\pi v_F}\int^{\pi}_{0} d\chi \cos^2\chi \delta\left(\dfrac{{\omega}^2}{{v_F^2 k}^2}- \cos^2\chi\right).
\end{align}
The principal value in the subsonic regime is found to simply be a constant by partial fraction expansion:
\begin{equation}
    \dfrac{k_F g^2}{4\pi^2 v_F} \mathcal{P} \int^{\pi}_{0} d\chi \dfrac{\cos^2\chi}{\dfrac{{\omega}^2}{{v_F^2 k}^2}- \cos^2\chi}=-\dfrac{k_F g^2}{4\pi v_F}.
\end{equation}
The result is thus solely a (negative) contribution to the squared mass of the boson:
\begin{equation}
    m^2=m_0^2-\dfrac{k_F g^2}{4\pi v_F}.
\end{equation}
which can be cancelled by renormalizing the boson's bare mass
\begin{equation}
    m_0^2=\dfrac{k_F g^2}{4\pi v_F}.
\end{equation}
The lesson is that the critical point should be defined in a way that takes the interaction with the Fermi surface into account. Ignoring the mass renormalization, one is thus left with the contribution from the pole, leading to
\begin{equation}\label{spherical}
     \Pi\left(\omega,\boldsymbol{k}\right)=-i\dfrac{k_F g^2}{4\pi v_F} \left({\dfrac{v_F^2 k^2}{\omega^2}-1}\right)^{-\frac{1}{2}},
\end{equation}
so the interesting contribution comes from a single resonant patch, which always exists in the subsonic regime\textemdash as advertized. This is a standard result also found in e.g. \cite{lawler} and \cite{sohn}. It is plotted in figure \ref{fig:self-energy}. Note that the result only depends on the ratio of $\omega$ and $k$\textemdash the only energy scale is the momentum. The low-energy and low-momentum limit thus do not commute\textemdash the low energy limit is the high momentum limit. Of course, the entire validity of the result depends on the momentum's being much lower than $k_F$, but both the ``low momentum'' and ``high momentum'' limits as used in this context are consistent with this requirement.  In the low energy limit, the result coincides with the famous $\frac{\left|\omega\right|}{\left|\boldsymbol{k}\right|}$ behavior of the boson's self energy, obtained by large-$N$ methods in \cite{polchinski}\cite{Sachdev}.

The formal result \eqref{spherical} also turns out to analytically continue to the supersonic regime, $\omega> v_F k$\textemdash except that since \eqref{spherical}  turns from purely imaginary to purely real, its interpretation changes from pure decay into a pure correction to the dispersion. This dramatic change is to be expected because there are no electron-hole pairs to which the boson can decay in the supersonic regime\textemdash moving faster than the speed of the Fermi surface, it behaves like a shock wave that cannot be scattered by the medium. In the ``low momentum'' limit (or ``high energy'' limit, which can actually involve arbitrarily small $\omega$!), the correction to the dispersion just adds a mass, which is nothing but the ``Schwinger mass'' calculated exactly in \cite{goldman} for the critical Fermi surface in the same low-momentum limit using the chiral anomaly. Again, the analogy with Schwinger's exact solution of 1+1-dimensional QED is striking\cite{schwinger}.

When the boson moves at exactly the Fermi velocity, $\omega=v_F k$, there is a pole in the self-energy. This pole is analogous to the ${\left(\omega-v_F k\right)^{-\alpha}}$ pole found in the one-dimensional Luttinger liquid, and can be viewed as a resonance between the boson and the Fermi surface's density waves. However, the behavior of the pole becomes more surprising in the non-isotropic case, and this is the novel result that was mentioned above.

In the non-isotropic case, the principal value is no longer a mere constant, but the imaginary part of the self-energy can still be easily calculated:

\begin{align}\label{pi-vf}
    \nonumber \Im \Pi(\omega,\boldsymbol{k})&=- \dfrac{ g^2}{\pi}\int^{\pi}_{0} d\chi \dfrac{k_F (\chi)}{v_F(\chi)} \delta\left(1-v_F(\chi)^2\frac{ k^2}{\omega^2} \cos^2\left(\theta-\chi\right)\right)\\
    &=-\dfrac{g^2}{4\pi}  \dfrac{k_F\left(\chi^{\star}\right)}{v_F\left(\chi^{\star}\right)} {\left|\sqrt{v_F^2\left(\chi^{\star}\right)\dfrac{ k^2}{\omega^2}-1}-\left|\dfrac{{v^\prime_{F}}\left(\chi^{\star}\right)}{v_F\left(\chi^{\star}\right)}\right|\right|}^{-1},
\end{align}
where $\chi^{\star}$ is now defined implicitly as the solution of the resonance equation
\begin{equation}\label{imp}
    v_F \left(\chi^{\star}\right)\cos\left(\theta-\chi^{\star}\right)=\pm\dfrac{\left|\omega\right|}{\left|\boldsymbol{k}\right|}.
\end{equation}

\begin{figure}
    \centering
        \captionsetup{justification=centerlast}
    \includegraphics[width=8.5 cm]{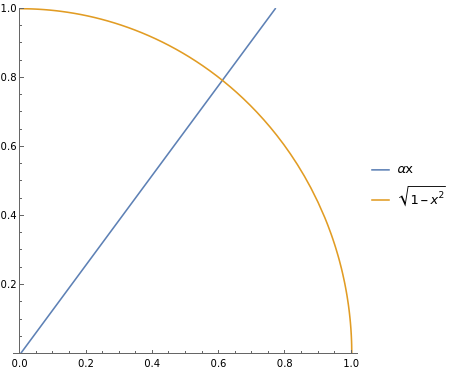}
    \caption{Any positive line passing through the origin intersects with a quarter circle centered around the origin. Hence, the self-energy \eqref{pi-vf} has a pole for any $\chi^{\star}$.}
    \label{fig:pole}
\end{figure}

In general, there is no pole when the boson moves at the Fermi velocity of the patch, $\omega=v_F\left(\chi^{\star}\right) k$. Nevertheless, for any given momentum, \eqref{pi-vf} \textit{does} diverge at \textit{some} subsonic frequency, i.e., lower than $v_F\left(\chi^{\star}\right) k$. Observe that the denominator vanishes if
\begin{equation}\label{pole}
    \sqrt{1-\dfrac{\omega^2}{v_F^2\left(\chi^{\star}\right)k^2}}+\dfrac{{v^\prime_{F}}\left(\chi^{\star}\right)}{v_F\left(\chi^{\star}\right)}\dfrac{\left|\omega\right|}{v_F\left(\chi^{\star}\right)\left|\boldsymbol{k}\right|}=0.
\end{equation}
Note that the resonance condition \eqref{imp} has a solution for any (subsonic) combination of $\omega$, $\left|\boldsymbol{k}\right|$ \textit{and} $\chi^{\star}$, provided that one sets $\theta$ appropriately\textemdash thus, one can regard $\chi^{\star}$ as some fixed, given constant. For given $\chi^{\star}$, equation \eqref{pole} describes the intersection between a quarter circle and a straight line:
\begin{align}
    \sqrt{1-x^2}+\dfrac{{v^\prime_{F}}}{v_F }x=0 && \text{where  }\text{ }0\leq x \leq 1,
\end{align}
which has a unique solution:
\begin{equation}
    x=\dfrac{1}{\sqrt{1+\frac{v^{\prime 2}_{F}}{v_F^2}}}
\end{equation}
(see figure \ref{fig:pole}). This solution depends on the direction of the incoming boson's momentum through $\chi^{\star}$, so rather than a single pole, one finds different poles for bosons moving in different directions. 

While \textit{some} pole has survived the angular dependence, this pole is not quite \textit{the} pole obtained for a Fermi sphere. In the spherically symmetric case, the self-energy diverges as
\begin{equation}
    \Pi(\epsilon)\sim \dfrac{ k_F g^2}{v_F} {\dfrac{1}{\sqrt{2 \epsilon}}}
\end{equation}
where $\epsilon=\frac{\omega-v_F k}{v_F k}$. In equation \eqref{pi-vf}, by contrast, $\sqrt{\frac{v_F^2 k^2}{\omega^2}-1}$ is analytic at the pole, and so the self-energy diverges linearly:

\begin{equation}
    \Pi(\epsilon)\sim-\dfrac{i k_F g^2 \left|v^\prime_F\right|}{ \left(1+\frac{v^{\prime 2}_{F}}{v_F^2}\right)^{\frac{3}{2}} v_F^2 } {\dfrac{1}{\left|\epsilon\right|}}
\end{equation}
where $\epsilon=\frac{\omega-\omega_\text{pole}}{\omega_\text{pole}}$.

The pole in the self-energy has thus been shifted to subsonic frequencies. However, $\omega=v_F\left(\chi^{\star}\right) k$ remains the threshold above which no on-shell density waves can be excited (again, treating $\chi^{\star}$ as a fixed constant by choosing $\theta$ appropriately)\textemdash above it, the self-energy is purely real. Therefore, rather than diverging, the self-energy \textit{jumps} at $\omega=v_F k$\textemdash its imaginary part jumps from some finite value down to zero. To our knowledge, these results regarding the pole and the jump in the self-energy are new.

In a future publication, we will provide a simple explanation of the full functional dependence of the boson's self energy based on fundamental physical principles.

\section{Exact Fermion Green's Function}\label{section5}
\subsection{General Green's Function}\label{exact}
We now turn to the evaluation fermion Green's function. Lawler et al. calculated it for equal times and equal positions \cite{lawler}, assuming a quadrupolar Landau interaction between fermions at different patches (we are assuming no such coupling). However, the full Green's function has an interesting scale invariance missing from the equal-time and equal-position results, and is necessary in order to reproduce the standard $\omega^{2/3}$ self-energy, so we will discuss it at length.

In light of the derivation in Section \ref{section3} in which $\zeta$ arises as a phase field, and by analogy with the one-dimensional Luttinger liquid, one can identify the electron's field with an exponential of the chiral mode of the phase. This can be derived more rigorously by inserting powers of the electron field $\psi$ to the path integral, and repeating the derivation of Section \ref{section3} (i.e., shifting the chiral phase of the Dirac spinor $\Psi$ to cancel the interaction term and integrating the fermion field out). Note also that if the phase is a generator of shifts of the charge, then its exponential would shift the charge by one unit. 

The outgoing mode of the phase-field is obtained in the path integral formalism by projecting out the incoming mode:
\begin{equation}
    \zeta_{+}(\boldsymbol{k},\omega,\chi)=\left(1-\dfrac{\omega}{v_F k_{\perp}}\right) \zeta (\boldsymbol{k},\omega,\chi).
\end{equation}
In terms of it, the fermion field is given by
\begin{equation}\label{ferm-op}
    \psi\left(\boldsymbol{x},t,\chi\right)=\exp\left[i\zeta_{+} \left(\boldsymbol{x},t,\chi\right)\right],
\end{equation}
Using the rules of Gaussian integration, the fermion propagator is thus given by
\begin{align}\label{prop}
    \nonumber G_{+}(t,x)&=\left<\exp\left[i\zeta_{+} \left(\boldsymbol{x},t,\chi\right)\right] \exp\left[-i\zeta_{+} \left(\boldsymbol{0},0,\chi\right)\right]\right>\\
    &=\exp\left[\left<\zeta_{+} \left(\boldsymbol{x},t,\chi\right)\zeta_{+} \left(\boldsymbol{0},0,\chi\right)\right>\right].
\end{align}

The electron-hole pair propagator is given by the sum of the free propagator and an interacting correction related to the self energy. The reason is that an electron-hole pair can either emit or not emit a boson, and the contribution to the propagator in the latter case is nothing but a free propgagtor. Hence
\begin{equation}
    \left<\zeta_{+} \left(\boldsymbol{x},t,\chi\right)\zeta_{+} \left(\boldsymbol{0},0,\chi\right)\right>\equiv \Delta_{+}(\boldsymbol{x},t)=\Delta^{(0)}_{+}(\boldsymbol{x},t)+\Delta^{\prime}_{+}(\boldsymbol{x},t),
\end{equation}
so the Fermion propagator \eqref{prop} is a product
\begin{equation}
    G_{+}(\boldsymbol{x},t)= G^{(0)}_{+} (\boldsymbol{x},t) G^{\prime}_{+}(\boldsymbol{x},t).
\end{equation}

Unsurprisingly, $G^{(0)}_{+} (\boldsymbol{x},t)$ is essentially the propagator of a free, 1+1D massless fermion. However, some subtleties arise in its derivation due to the highly singular nature of tangential (non-)dispersion. Since this is a rather subtle discussion, and since the free fermion problem is more standard and less interesting than the critical Fermi surface, we leave the derivation to Appendix \ref{appendix1}. Here, we just state the final answer:
\begin{align}\label{free-prop}
 G^{(0)}_{+}\left(\boldsymbol{x},t\right) =\delta\left(x_{\parallel}\right) \dfrac{1}{x_{\perp}-v_F t+i\epsilon \text{sign}\left(x_{\perp}\right)}.
\end{align}

\begin{figure}
    \centering
        \captionsetup{justification=centerlast}
    \includegraphics[width=10cm]{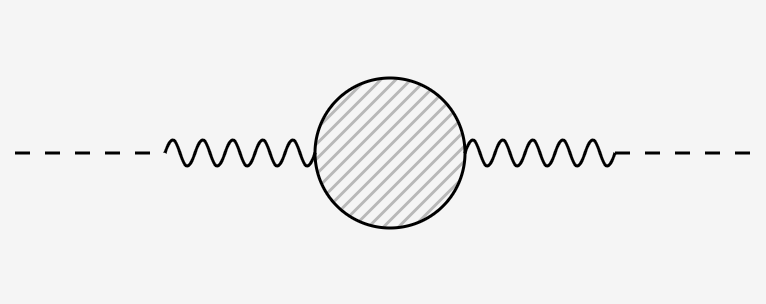}
    \caption{The interacting correction to the density wave propagator is as follows: A density wave can scatter into a boson, which can do whatever it can do, but in the end has to scatter back into a density wave.}
    \label{fig:density-prop}
\end{figure}

The interacting correction to the electron-hole propagator $\Delta^{\prime}\left(\boldsymbol{x},t\right)$ is just the amplitude that the electron-hole pair scatter into a \textit{dressed} boson (i.e., one that emit an electron-hole pair any number of times) and that the boson then scatters again into an electron-hole pair, as shown in figure  \ref{fig:density-prop}. This amplitude is proportional the product of two chiral density wave propagators and the full boson propagator
\begin{align}\label{inter-prop}
    \Delta^{\prime}_{+} (\omega,\boldsymbol{k})&=\dfrac{g^2 k_F^{2}}{16 \pi^2} k_{\perp}^2 \Delta^{(0)}_{+}(\omega,\boldsymbol{k}) D(\omega,\boldsymbol{k}) \Delta^{(0)}_{+}(\omega,\boldsymbol{k}) \nonumber \\
    &\approxeq  -\dfrac{4\pi^2 g^2}{\left(\omega-v_F k_{\perp}\right)^2} \dfrac{1}{k^2_{\parallel}+i\dfrac{g^2 k_F}{4 \pi v_F}\dfrac{\left|\omega\right|}{\left|k_{\parallel}\right|}}.
\end{align}
Here, we have taken a low energy limit: we have thus neglected the $\omega^2$ term in the boson's propagator, which is subdominant at low energies compared to the $\frac{\left|\omega\right|}{\left|\boldsymbol{k}\right|}$ term. We have also approximated $\left|\boldsymbol{k}\right|$ as $\left|k_{\parallel}\right|$, which is exact at low energy (where tangential scattering dominates, since it costs zero energy). 

Equation \eqref{inter-prop} is a product of a term with no $k_{\parallel}$-dependence and a term with no $k_{\perp}$-dependence. Transforming into position space is thus rather simple: the 2D spatial Fourier transform factorizes into a product of two 1D Fourier transforms. The normal Fourier transform in particular is straightfowrad, and it is similar to the Fourier transform of a free particle's Green's function in classical mechanics
\begin{equation}\label{class-m}
    \int^{\infty}_{-\infty} \dfrac{dk_{\perp}}{2\pi}  \dfrac{e^{-i k_{\perp} x_{\perp}}}{\left(\omega-v_F k_{\perp}\right)^2}=-\dfrac{1}{2 v_F^2} \left|x_{\perp}\right| \exp\left(\dfrac{i}{v_F}\omega x_{\perp}\right),
\end{equation}
so the only dependence on $\omega$ is an operator that shifts time by $x_{\perp}/v_F$\textemdash which is to be expected since the electrons move outward at the Fermi velocity.

The tangential Fourier transform looks slightly more threatening. However, because of the $\delta\left(x_{\parallel}\right)$ from the free propagator, the only contribution to the fermion propagator will be from $x_{\parallel}=0$. The $x_{\parallel}=0$ Fourier transform can be computed by dimensional analysis, owing to the scale invariance of the boson propagator:
\begin{align}
    \int^{\infty}_{-\infty} \dfrac{dk_{\parallel}}{2\pi}\dfrac{1}{k_{\parallel}^2+i\dfrac{g^2 k_F}{4\pi v_F} \frac{\left|\omega\right|}{\left|k_{\parallel}\right|}}&=\left(\dfrac{g^2 k_F}{4 \pi v_F}\right)^{-1/3}\left|\omega\right|^{-1/3} \int^{\infty}_{-\infty} \dfrac{ds}{2\pi}\dfrac{1}{s^2+\frac{i}{\left|s\right|}}\nonumber\\
    &=\dfrac{2}{3\sqrt{3}} \left(\dfrac{g^2 k_F}{\pi v_F}\right)^{-1/3}  e^{-i\pi/6}\left|\omega\right|^{-1/3},
\end{align}
and so the only dependence on $\omega$ is a power law which will Fourier transform into another power law. This power law would diverge at small comoving distance $x_{\perp}-v_F t$. Consequently, the pole of \eqref{free-prop} at $x_{\perp}=v_F t$ is wiped out\textemdash yet it is precisely the residue from this pole that determines the free off-shell propagator. However, the divergence of the Fourier transform owes to the contribution from indefinitely high frequencies, and by setting an appropriate high frequency cutoff $\Gamma$ no divergence arises. With the cutoff, one obtains
\begin{align}
\nonumber    \Delta^{\prime}_{+}\left(x_{\perp},x_{\parallel}=0,t\right)&=-\beta e^{-i\pi/6} \left|x_{\perp}\right| \int^{\Gamma}_{-\Gamma} \dfrac{d\omega}{2\pi} \left|\omega\right|^{-1/3} \exp\left[\dfrac{i}{v_F}\omega \left(x_{\perp}-v_F t\right)\right]\\&= -\beta e^{-i\pi/6} {\left|x_{\perp}\right|} H\left(\Gamma \left|x_{\perp}-v_F t\right|\right),
\end{align}
where
\begin{equation}
    \beta= \Gamma^{2/3}\dfrac{16\pi^2 }{3\sqrt{3}}\left(\dfrac{\pi g^4}{k_F v_F^4}\right)^{1/3}, 
\end{equation}
and
\begin{align}
\nonumber H(z)&=\dfrac{1}{\pi}\int^{1}_{0} ds\text { }s^{-1/3} \cos\left(z s\right)\\
&=  \dfrac{3}{2\pi} \prescript{}{1}F_2\left(\dfrac{1}{3};\dfrac{1}{2},\dfrac{4}{3};-\dfrac{1}{4} z^2\right),
\end{align}
with $F$ the generalized hypergeometric function. The important properties of the complicated-looking $H$ function are its two extreme limits: At comoving distances much smaller than the cutoff, it is just a finite constant insensitive to the distance
\begin{equation}
    H(0)=\dfrac{1}{\pi} \int^{1}_{0} ds\text{ } s^{-1/3}=\dfrac{3}{2\pi}
\end{equation}
which will govern the residue from the $x_{\perp}=v_F t$ pole. At comoving distances much larger than the cutoff, the result is a power law insensitive to the cutoff
\begin{align}
    H(z)&\overset{z\rightarrow\infty}{\sim} \dfrac{1}{\pi}\int^{\infty}_{0} ds\text{ } s^{-1/3} \cos\left(z s\right)\nonumber\\
    &=\dfrac{\Gamma\left(2/3\right)}{2\pi} \left|z\right|^{-2/3}
\end{align}
which will govern the behavior of the propagator far from the pole.

All in all, one finds 
\begin{align}\label{elect-prop-uncut}
    \nonumber G_{+}(\boldsymbol{x},t)=&G^{(0)}_{+}(\boldsymbol{x},t) \exp\left( \Delta^{\prime}_{+}\left(x_{\perp},x_{\parallel}=0,t\right)\right)\\
    \nonumber=&\dfrac{\delta\left(x_{\parallel}\right)}{x_{\perp}-v_F t +i\epsilon \text{sign}\left(x_{\perp}\right)} \exp\left[-\beta e^{-i\pi/6} {\left|x_{\perp}\right|} H\left(\Gamma \left|x_{\perp}-v_F t\right|\right)\right]\\
    =&\nonumber-i\pi \text{sign}\left(x_{\perp}\right) \delta\left(x_\parallel\right)\delta\left(x_{\perp}-v_F t\right) \exp\left(-\frac{3}{2\pi}\beta e^{-i\pi/6} {\left|x_{\perp}\right|}\right)\\
    &+\mathcal{P} \dfrac{\delta\left(x_{\parallel}\right)}{x_{\perp}-v_F t} \exp\left[-\beta e^{-i\pi/6} {\left|x_{\perp}\right|} H\left(\Gamma \left|x_{\perp}-v_F t\right|\right)\right].
\end{align}

Equation \eqref{elect-prop-uncut} is the exact fermion Green's function. While it is valid under the assumption that the fermion emits bosons with $\omega\ll v_F k$, the propagator's Fourier transform is clearly no longer a function of $\frac{\omega}{v_F k}$ alone, and so $v_F k$ is no longer the only energy scale\textemdash there will be a variety of different behaviors at different energies. The celebrated $\omega^{2/3}$ self-energy is obtained in the special limit of low energies, in which the Green's function simplifies considerably. 

\subsection{Long Comoving Distance Green's Function}
The low energy limit is related, but not equivalent, to the asymptotic behavior of the real-space propagator far from its pole, $x-v_F t\rightarrow \infty$. The difference between the two limits is that the delta-singularity contributes to all energy scales indiscriminately. Nevertheless, the limit of large $x-v_F t$ already illustrates many of the essential properties of the less obvious low energy limit, and is interesting in its own right. Hence, we will start by discussing it. 

For large $x-v_F t$, one may ignore the cutoff and the pole, and the Green's function can be written as
\begin{align}\label{elect-prop-as}
     G_{+}(\boldsymbol{x},t)\approxeq \dfrac{\delta\left(x_{\parallel}\right)}{x_{\perp}-v_F t} \exp\left(-\gamma e^{-i\pi/6} \dfrac{\left|x_{\perp}\right|} {\left|x_{\perp}-v_F t\right|^{2/3}}\right),
\end{align}
where we have defined
\begin{equation}
    \gamma=\dfrac{4 \Gamma(2/3) }{3\sqrt{3}}\left(\dfrac{\pi^4 g^4}{k_F v_F^2}\right)^{1/3}. 
\end{equation}
The asymptotic Green's function \eqref{elect-prop-as} is consistent with the equal-time result obtained by Lawler et al. in \cite{lawler}, which is of the form $\frac{1}{|x|}\exp\left(-C\left|x\right|^{1/3}\right)$ but not with the equal-position result, which is $\frac{1}{\left|t\right|}\exp\left(-D\left|t\right|^{2/3}\right)$\textemdash instead, their equal position result behaves like our result after a long time \textit{and at a finite but smaller distance}, $x\ll  v_F t$. This may be a result of the fact they explicitly impose a cutoff on the normal momentum in their calculation of the Fourier integrals, breaking the scale invariance of the result. In this case, our result should be valid for positions exceeding the cutoff, but not at strictly equal positions.

However, at positions exceeding the cutoff and at arbitrary times, the result reveals a remarkable nature of the Green's function not apparent from the equal-position and at equal-time results. It is similar to the diffusive Green's function
\begin{equation}
    G_{\text{diffusion}}(x,t)=\dfrac{1}{\sqrt{4\pi Dt}}\exp\left(-\dfrac{x^2}{4Dt}\right),
\end{equation}
with the two obvious differences being that in \eqref{elect-prop-as} the exponents of $x$ and $t$ are anomalous, and that the null coordinate $x_{\perp}-v_F t$ plays the role of time. Just like its diffusive counterpart, \eqref{elect-prop-as} has scale invariance\textemdash an emergent low-energy scale invariance not apparent in the microscopic action \eqref{weyl-lin}. This scale invariance mixes space and time:
\begin{align}
    x_{\perp}&\rightarrow \alpha x \nonumber\\
    x_{\perp}-v_F t&\rightarrow \alpha^{3/2} \left(x_{\perp}-v_F t\right) \nonumber \\
    G_{+}(\boldsymbol{x},t)&\rightarrow \alpha^{-3/2} G_{+}(\boldsymbol{x},t).
\end{align}
One can easily convert this to momentum space, leading to a scaling of the frequency $\omega$ and the null momentum $\omega-v_F k$:
\begin{align}\label{scaling-momentum}
    \omega-v_F k_{\perp}&\rightarrow \alpha^{-1} \left(\omega-v_F k_{\perp}\right) \nonumber\\
    \omega &\rightarrow \alpha^{-3/2} \omega \nonumber \\
    G_{+}(\boldsymbol{k},\omega)&\rightarrow \alpha G_{+}(\boldsymbol{k},\omega).
\end{align}
The general form of $G_{+}(\boldsymbol{k},\omega)$ consistent with this scaling, with parity and with time reversal, is 
\begin{equation}
    G_{+}(\boldsymbol{k},\omega)=\dfrac{A\left(\frac{\text{sign}(\omega)\omega^{2/3}}{\omega-v_F k_{\perp}}\right)}{\omega-v_F k_{\perp} +B\left(\frac{\text{sign}(\omega)\omega^{2/3}}{\omega-v_F k_{\perp}}\right) \text{sign}(\omega)\omega^{2/3}}.
\end{equation}
The two extreme limits of this expression, are, similarly to the bosonic case, low energies and high null momenta $\omega-v_F k_{\perp}\gg \left(\frac{g^4 v_F}{k_F }\right)^{1/3} \omega^{2/3}$, and high energies and low momenta $\omega-v_F k_{\perp}\ll \left(\frac{g^4 v_F}{k_F}\right)^{1/3} \omega^{2/3}$. If the coefficients $A$ and $B$ were not singular at zero energy, then the leading order low energy propagator would be the standard result:

\begin{equation}\label{dominant}
        G_{+}(\boldsymbol{k},\omega)=\dfrac{A_0}{\omega-v_F k_{\perp} +B_0 \text{sign}(\omega) \omega^{2/3}}.
\end{equation}
However, because of the antisymmetry of \eqref{elect-prop} in $x_{\perp}-v_F t$, the zero frequency Fourier transform actually vanishes:
\begin{equation}
    G_{+}(\omega=0,\boldsymbol{x})=\int d^2 x\dfrac{d\left(x_{\perp}-v_F t\right)}{v_F} e^{i\boldsymbol{k}\cdot\boldsymbol{x}} G_{+}(x_{\perp}-v_F t,\boldsymbol{x})=0,
\end{equation}
and in Appendix \ref{appendix2} it is shown that at low energies the Fourier transform scales as $\frac{\omega^{3/2}}{\left(\omega-v_F k_{\perp}\right)^2}$. This apparent discrepancy with the standard result arises because the long (comoving-)distance limit is not equivalent to the low energy limit\textemdash the low energy limit depends crucially on the residue from the $x=v_F t$ pole. Such is the reason why it was necessary to explicitly impose the high-energy cutoff in Subsection \ref{exact}.

\subsection{Low Energy Green's Function}
It turns out that the standard $\omega^{2/3}$ result is recovered in the strongest possible low-energy limit, in which the null momentum $k_{\perp}-\omega/v_F$ is much greater not only than the momentum scale constructed from the energy $\omega$, but than the one constructed from \textit{the high-energy cutoff} $\Gamma$:

\begin{equation}\label{strong-ir-limit}
    \omega-v_F k_{\perp}\gg \left(\dfrac{g^4 v_F}{k_F}\right)^{1/3} \Gamma^{2/3}.
\end{equation}
While it may appear odd that the \textit{high}-energy cutoff is to be treated as a \textit{small} parameter, observe that it is suppressed by a factor of $k_F^{-1/3}$, and $k_F$ is taken to infinity in the double scaling limit. Hence, \eqref{strong-ir-limit} does not require that the cutoff $\Gamma$ be small or even finite, but merely that $k_F$ be sufficiently large compared to $\frac{\Gamma^{2}}{\left(\omega-v_F k_{\perp}\right)^3}$. Of course, this is perfectly consistent with the largeness of the cutoff \textit{compared to the energy}, $\Gamma \gg \omega$\textemdash in fact, it will be shown that the undetermined cutoff $\Gamma$ drops out of the leading order result.

Recall that the full propagator is 
\begin{align}\label{elect-prop-uncut2}
    G_{+}(\boldsymbol{x},t)=&\nonumber-i\pi \text{sign}\left(x_{\perp}\right) \delta\left(x_\parallel\right)\delta\left(x_{\perp}-v_F t\right) \exp\left(-\frac{3}{2\pi}\beta e^{-i\pi/6} {\left|x_{\perp}\right|}\right) \nonumber \\
    &+\mathcal{P} \dfrac{\delta\left(x_{\parallel}\right)}{x_{\perp}-v_F t} \exp\left[-\beta e^{-i\pi/6} {\left|x_{\perp}\right|} H\left(\Gamma \left|x_{\perp}-v_F t\right|\right)\right].
\end{align}
In the high-momentum, low energy limit \eqref{strong-ir-limit}, the dominant contribution to the residue from the pole will be from the jump of the sign function, so one can approximate
\begin{equation}
     \exp\left(-\frac{3}{2\pi}\beta e^{-i\pi/6} {\left|x_{\perp}\right|}\right)\approxeq 1.
\end{equation}
Corrections will be small since the exponent decays very slowly ($\beta \ll v_F k_{\perp}-\omega$), and it can be readily shown that they  scale as $\frac{\Gamma^{4/3}}{\left(\omega -v_F k_{\perp}\right)^2}$. 

Furthermore, since $\Gamma \gg \omega$, the dominant contribution to the principal value will be from comoving distances $x-v_F t$ far exceeding the inverse energy cutoff, so one can approximate\footnote{Note that it would be going too far to approximate the exponent as unity, as was done in the case of the residue from the pole. Doing so would erase all dependence of the principal value on $x_{\perp}$, thereby turning it into a delta-function of the null momentum\textemdash too crude an approximation if one wishes to understand the energy and momentum-dependence of the propagator off-shell.}
\begin{equation}
     \exp\left[-\beta e^{-i\pi/6} {\left|x_{\perp}\right|} H\left(\Gamma \left|x_{\perp}-v_F t\right|\right)\right]\approxeq \exp\left(-\gamma e^{-i\pi/6} \dfrac{\left|x_{\perp}\right|} {\left|x_{\perp}-v_F t\right|^{2/3}}\right).
\end{equation}
Corrections will be from the small interval $-v_F \Gamma^{-1}\lesssim x-v_F t\lesssim  v_F \Gamma^{-1}$, and their effect would likewise be small, as can be easily assessed using the antisymmetry of the principal value term
\begin{align}
    \int^{v_F\Gamma^{-1}}_{-v_F\Gamma^{-1}} du\text{ } e^{i\frac{\omega}{v_F} u} \dfrac{\exp\left[-\beta e^{-i\pi/6} {\left|x_{\perp}\right|} H\left(\Gamma \left|u\right|\right)\right]}{u} &\sim \int^{v_F\Gamma^{-1}}_{-v_F\Gamma^{-1}} du \dfrac{\sin\left[\frac{\omega}{v_F}u\right]}{u} \nonumber\\
    &=  \int^{\Gamma/v_F}_{-\Gamma/v_F} du  \dfrac{\omega}{v_F}+\mathcal{O}\left(\dfrac{\omega^2}{ \Gamma^{2}}\right)\nonumber\\
    &\sim\dfrac{\omega}{\Gamma},
\end{align}
where $u\equiv x-v_F t$.

The total low-energy propagator is thus identical to the long-distance propagator \eqref{elect-prop} up to a simple residue from the pole
\begin{align}\label{elect-prop}
     G_{+}(\boldsymbol{x},t)\approxeq& -i\pi \delta\left(x_{\perp}-v_F t\right)\delta\left(x_{\parallel}\right)  \text{sign}\left(x_{\perp}\right)\nonumber\\&+ \dfrac{\delta\left(x_{\parallel}\right)}{x_{\perp}-v_F t} \exp\left(-\gamma e^{-i\pi/6} \dfrac{\left|x_{\perp}\right|} {\left|x_{\perp}-v_F t\right|^{2/3}}\right).
\end{align}
As claimed before and shown in Appendix \ref{appendix2}, the long-distance propagator Fourier transforms to 
\begin{equation}
     \dfrac{\delta\left(x_{\parallel}\right)}{x_{\perp}-v_F t} \exp\left(-\gamma e^{-i\pi/6} \dfrac{\left|x_{\perp}\right|} {\left|x_{\perp}-v_F t\right|^{2/3}}\right)\rightarrow -B \text{sign}(\omega) \dfrac{\omega^{2/3}}{\left(\omega-v_F k_{\perp}\right)^2}+\mathcal{O}\left({\dfrac{\omega^2}{k^4}}\right),
\end{equation}
where
\begin{equation}
     B\equiv\dfrac{4}{\sqrt{3}} e^{-i\pi/3} \left(\dfrac{\pi^{4} g^{4}}{k_F v_F^4}\right).
\end{equation}

The Fourier transform of the residue from the pole can also be easily computed, and gives
\begin{equation}
     -i\pi \delta\left(x_{\perp}-v_F t\right)\delta\left(x_{\parallel}\right)  \text{sign}\left(x_{\perp}\right) \rightarrow \dfrac{1}{\omega-v_F k_{\perp}}.
\end{equation}
Combining, one finds
\begin{align}
    G_{+}\left(\omega,\boldsymbol{k}\right)&=\dfrac{1}{\omega-v_F k_{\perp}} -B \text{sign}(\omega) \dfrac{\omega^{2/3}}{\left(\omega-v_F k_{\perp}\right)^2}+\mathcal{O}\left({\dfrac{\omega^2}{k^4}}\right)\nonumber\\
    &=\dfrac{1}{\omega-v_F k_{\perp}+B\text{sign}(\omega) \omega^{2/3}}+\mathcal{O}\left({\dfrac{\omega^2}{k^4}}\right),
\end{align}
which is the standard and celebrated result obtained in the large-$N$ theory\cite{polchinski}. 

All the important of the large-$N$ theory have thus been reproduced without any large-$N$ approximation, and without any tangential term in the electron dispersion. In what follows, we will assess whether ignoring the tangential term has truly been justified.

\section{Discussion: Can the Dispersion Really Be Linearized?}\label{discussion}
\subsection{Irrelevance of Quadratic Dispersion in Multi-Patch Theories}\label{irrelevance}
Presumably, our assertion most prone to the reader's suspicion is that the the tangential dispersion is irrelevant at low momenta, as all the rest follows from it. In particular, one might think that since $k_{\perp}$ and $k_{\parallel}$ are independent variables, it is conceivable that the quadratic order of the latter is comparable to the linear order of the former. In fact, at low energies, it seems like it \textit{must} be the case, because the normal momentum will be essentially zero, so \textit{all} the scattering will be in the tangential direction. 

The crucial point is that since the momentum and the energy of an electron-hole pair are transferred to it by the incoming boson, the negligibly of the tangential dispersion is nothing but a condition on the incoming boson's state. To explicitly assess whether the condition is reasonable, consider for simplicity an isotropic dispersion
\begin{equation}
    \varepsilon\left(\boldsymbol{p}\right)=\varepsilon\left(\left|\boldsymbol{p}\right|\right).
\end{equation}
Now, using 
\begin{equation}\label{relative-to-patch1}
    \left|\boldsymbol{p}\right|=\sqrt{\left(k_{\perp}+k_F\right)^2+k_{\parallel}^2},
\end{equation}
one can expand the dispersion while keeping the leading tangential term to obtain
\begin{equation}
     \varepsilon(\boldsymbol{k})= v_F k_{\perp}+\dfrac{v_F}{2k_F}k_{\parallel}^2+\mathcal{O}\left(k^3\right).
\end{equation}
At low enough energies, one can assume that the entire momentum is tangential, $k_{\parallel}^2\approxeq k^2$. The ratio between the tangential term and the normal term is then the dimensionless parameter
\begin{equation}
    \eta \equiv \dfrac{k^2}{k_F k_{\perp}}=\dfrac{1}{\dfrac{k_F}{k}\dfrac{\omega}{v_F k}-1}.
\end{equation}
It is seen that $\eta$ is small insofar as
\begin{equation}\label{condition}
    \dfrac{\omega}{v_F k}\gg \dfrac{k}{k_F}.
\end{equation}
This definitely holds when the phase velocity of the boson is at least of the order of the Fermi velocity, in which case $\frac{\omega}{v_F k}$ is not a small parameter. But it also holds for infinitesimal $\frac{\omega}{v_F k}$, provided only that the other infinitesimal parameter $\frac{k}{k_F}$ is even smaller. This defines a consistent IR limit in which $k\rightarrow0$ and $\omega\rightarrow 0$ in such a way that there is the \textit{hierarchy of dimensionless parameters} shown in table \ref{tab:hierarchy}. Differently put, $\omega$ should be bounded between a linear function of $k$ and a quadratic function of $k$.
\begin{table}[]
    \centering
    \captionsetup{justification=centerlast}
    \scalebox{1.5}{\begin{tabular}{c} 
         1 \\
          $\downarrow$\\
          $\dfrac{\omega}{v_F k}$\\
         $\downarrow$\\
         $\dfrac{k}{k_F}$\\
    \end{tabular}}
    \caption{Hierarchy of dimensionless parameters in the limit under which: (1) normal dispersion dominates; (2) the subsonic low-energy limit of Section \ref{section4} holds.}
    \label{tab:hierarchy}
\end{table}

Observe that the statement that normal dispersion dominates over tangential dispersion does \textit{not} assume the normal \textit{momentum} dominates over the tangential \textit{momentum} (which is clearly false at low energy). The validity of the limit depends not on the smallness of $k_{\parallel}$ relative to $k_{\perp}$, but instead on the smallness of \textit{both} relative to the Fermi momentum $k_F$.

This establishes that the linearized calculation of the boson self-energy is valid at low energies, as it does not couple different frequencies and momenta, so one can freely choose $\omega$ and $\boldsymbol{k}$ consistent with the condition \eqref{condition}. Yet the calculation of the electron self-energy in Section \ref{section5} involved a sum over \textit{all} energies and all momenta. The validity of the linearized limit thus depends on whether the dominant contribution to the sums satisfies it. In \cite{chubukov}, it is pointed out that the peak of the boson propagator is at momentum of the order of $\left(\frac{g^2 k_F}{v_F} \omega\right)^{1/3}$ (see \eqref{inter-prop}), so that at low energies the typical tangential energy exceeds the typical normal energy, suggesting that the former dominates.

However, this argument ignores the \textit{spread} around those typical values\textemdash if the spread is much greater than the peak, then the precise location of the peak is washed out. For the free fermion propagator, this is certainly the case: it is just a scale-free power law around $\omega=v_F k_{\perp}$, and shifting $\omega$ down would merely shift the pole in $k_{\perp}$. For the boson propgator, there is a momentum scale $\left(\frac{g^2 k_F}{v_F} \omega\right)^{1/3}$ characterizing the sharpness of the peak, and the propagator (viewed as a distribution function) can be normalized. Shifting $\omega$ down would not merely shift the location of the peak, but also make it increasingly sharper. This suggests that in spite of the smaller ``typical value'' the tangential dispersion, due to its equally small spread the answer is dominated by regions of phase space where the normal dispersion dominates, and eq. \eqref{condition} is satisfied. 

The counter-argument might seem suspicious in light of the long, power-law tail of the boson propagator at large momenta. It turns out to work nonetheless\textemdash the point is that at low energies, the quadratic dispersion $\beta k^2_{\parallel}$ would vary infinitely slowly compared to $D(\omega,k_{\parallel})$ around the latter's sharp peak, so the former can just be approximated as zero. This is confirmed by a calculation. Chubukov et al. argue \cite{chubukov} that the quadratic dispersion is relevant for the evaluation of the equal-time and equal position propagators that had been computed by Lawler et al.\cite{lawler} However, the low energy limit is defined by the condition $\omega-v_F k_{\perp}\gg \left(\frac{g^4 v_F}{k_F}\right)^{1/3} \omega^{2/3}$\textemdash or (aside from the $x_{\perp}=v_F t$ pole), $x_{\perp}\ll \left(\frac{g^4 v_F}{k_F}\right)^{1/3} \left(x_{\perp}-v_F t\right)^{2/3}$, while the double scaling limit requires $x_{\perp}\gg k_F$. The former is clearly not satisfied for the equal time propagator, while the latter is clearly not satisfied for the equal position propagator. Instead, it is the \textit{long-comoving distance and finite $x_{\perp}$} propagator (as well as the $x_{\perp}=v_F t$ pole) that governs the low-energy behavior of $G(\omega,\boldsymbol{k})$, and leads to the $\omega^{2/3}$ self-energy.

In Appendix \ref{appendix3}, we add quadratic dispersion in heuristic way suggested by \cite{chubukov}, but focus on the long comoving distance limit rather than the equal time or equal position limits. We show that corrections to the fermion propagator are subodminant. 

\subsection{Relevance of $N$th Order Dispersion in Single-Patch Theories}
At first glance, the above might seem to contradict the finding by many authors\cite{Sachdev} \cite{Sachdev2} \cite{mross} that under the renormalization group, the quadratic, tangential term in the dispersion is relevant. The reason why there is no contradiction is that these authors were considering a different class of models than our own, in which only a single patch of the Fermi surface, and possibly its antipode, are taken into account. Within such models, it is hardly surprising that the quadratic term in the dispersion is relevant\textemdash after all, without it one would be unable to derive the correct Landau damping\footnote{Incidentally, the Landau damping is the \textit{only} step in the one-loop calculation that depends on the quadratic dispersion\textemdash the one loop electron self-energy is independent of the dispersion so long as one can derive the correct Landau damping.}. Indeed, the quadratic dispersion is the last trace of the curvature of the Fermi surface in single patch theories, and discarding it amounts to ignoring the curvature.

However, in linearized multi-patch theories, the curvature is taken into account through the fact that different patches move in different directions, and the correct Landau damping is reproduced. 
To compare single-patch and multi-patch theories, one should express them in the same language. This can be done by rewriting the entire multi-patch action \eqref{weyl-lin} \textit{relative to a specific, fixed patch of $\boldsymbol{p}=\left(k_F,0\right)$}\textemdash where recall that $\boldsymbol{p}$ is the absolute momentum, to be distinguished from the relative momentum $\boldsymbol{k}$. To this end, define the momentum relative to a specific patch 
\begin{equation}
    \boldsymbol{q}=\left(p_x-k_F,p_y\right).
\end{equation}
In the continuum limit, the multi-patch model should converge to the linearized dispersion (again, assuming spherical symmetry for simplicity)
\begin{align}
    \varepsilon(\boldsymbol{p})= v_F  \left(\left|\boldsymbol{p}\right|-k_F \right)
\end{align}
In terms of the single-patch relative momentum $\boldsymbol{q}$, one can rewrite this as
\begin{equation}\label{relative-to-patch2}
    \varepsilon(\boldsymbol{q})=v_F \sqrt{\left(q_x+k_F\right)^2+q_y^2}-v_F k_F.
\end{equation}
Note that in spite the formal similarity to \eqref{relative-to-patch1}, \eqref{relative-to-patch2} has an entirely different interpretation: $q_x$ is and $q_y$ are the $x$ and $y$ components of momentum relative to a specific Fermi point with momentum in the $x$-direction, while $k_\perp$ and $k_\parallel$ are the components of the momentum relative to the \textit{nearest} patch, locally normal and tangential to the Fermi surface, respectively. The implication is that contrary to \eqref{relative-to-patch1}, at low energies, there is no reason to suppose that $q_x$ and $q_y$ are small\textemdash a patch different from the ``reference patch'' of the theory would have $\boldsymbol{q}$ of the order of $k_F$. Hence, to recover the full multi-patch theory one should keep track of all orders of $q_x$ and $q_y$:
\begin{equation}
    \varepsilon(\boldsymbol{q})=v_F \left[q_x+\dfrac{q_y^2}{2k_F}-\dfrac{q_y ^2 q_x}{2k_F^2}+...\right].
\end{equation}

While $\boldsymbol{q}$ need not be small, one might nevertheless hope that similarly to standard relativistic field theory, under the renormalization group the expansion truncates, and that high order terms are irrelevant at low energies. Indeed, when keeping only the quadratic term, the one loop results for the boson and fermion Green's functions precisely agree with those obtained in the multi-patch theory\cite{polchinski}. However, higher loop corrections diverge at low energies in the quadratic single-patch theory\cite{sung-sik-lee}, whereas in the multi-patch theory it has been shown that they \textit{vanish}. We are thus led to conjecture that while in multi-patch theories the quadratic dispersion is irrelevant, in single-patch theories one must go \textit{beyond} quadratic dispersion to correctly reproduce the higher loop results. Far from ignoring the curvature of the Fermi surface, the linearized multi-patch theory takes curvature into account \textit{more precisely} than quadratic multi-patch theories! The distinction between multi-patch theories and single-patch theories is summarized in table \ref{tab:single-vs-multi}.

\begin{table}[t]
    \centering
    \captionsetup{justification=centerlast}
\begin{tabular}{ | m{5cm} | m{4.3cm}| m{4.3cm} | } 
  \hline
   & \textbf{Single-patch theories} & \textbf{Multi-patch theories}
   \\ 
  \hline
  \textbf{Linear dispersion} & Relevant but insufficient at one loop & Relevant and sufficient at one loop; higher loops vanish\\ 
  \hline
  \textbf{Quadratic dispersion} & Relevant and sufficient at one loop; higher loops diverge & Irrelevant in the double-scaling limit and under the conditions of Section \ref{irrelevance}  \\
\hline  \textbf{Higher than quadratic dispersion} & Relevant at higher loops & Irrelevant
  \\ \hline
\end{tabular}
    \caption{Relevance of various powers in the dispersion in single-patch theories and in multi-patch theories.}
    \label{tab:single-vs-multi}
\end{table}

\section{Conclusion}
We have shown that there is a well-defined low-energy and low-momentum limit under which the simplest possible critical Fermi surface behaves simply, and can be solved exactly. The appeal of our approach is in that it is seemingly fully microscopic, and assumes no structure other than ordinary electrons and an ordinary bosonic  order parameter. Moreover, the absence of any inessential structure makes it particularly easy to interpret the mechanisms yielding some otherwise counter-intuitive results. In this paper we have established that this approach reproduces the essential results for critical Fermi surfaces obtained in the large-$N$ theory. 

Of course, there remains the question of whether our approach assumes \textit{enough} structure to capture \textit{all} the physics of critical Fermi surfaces\textemdash or whether it \textit{neglects} important effects. One important assumption that we have made is that the electrons are free save for the interaction with the critical boson. This does not seem too unreasonable given that the interaction with the boson is long-ranged and singular, while normal interactions are short-ranged and thus potentially subdominant. Yet any interaction between electrons would result in a coupling between patches, thus challenging the picture that the boson only couples to a single patch. Therefore, it remains an open question whether the coupling between different patches is relevant. Lawler et al. \cite{lawler} already considered the effect of a quadrupolar Landau interaction, but they obtained results consistent with our own. We have been trying to consider a different kind of an interaction between patches, and our initial results suggest that it may lead to new qualitative effects. This is a subject for a future publication.

In addition to the assumption that electron-hole pairs at different patches are decoupled, our approach assumes that the electron and the hole always reside at the patch. This is certainly a valid assumption when the electrons are only coupled to the critical boson\textemdash provided that the critical boson carries too small a momentum to excite an electron and a hole at two distant patches. Yet, a simple white noise disorder acts on all momenta indiscriminately, and thus excites electron-hole pairs at all momenta. The effect of disorder, or of some other momentum relaxation mechanism, is essential for understanding the celebrated linear-$T$ resistivity (for without scattering, electrons \textit{accelerate} in the presence of a constant electric field). It it for future research to study whether a ``gentler'' kind of disorder which acts on  each patch separately and decays at high momenta is sufficient to understand linear-$T$ resistivity\textemdash or whether the results of the model where electrons cannot hop between different patches can somehow be extended to the presence of disorder\textemdash or whether the model should be generalized to the case in which electron-hole pairs at two different patches can be excited (this can be done, e.g., by considering a bosonized field that depends on \textit{two} patches $\zeta\left(\chi,\chi^\prime\right)$). 

We make no claim that the effects of disorder and inter-patch-coupling are not essential in realistic experiments. Rather, we regard the simplest model as a ground on which to build in the future\textemdash as a zeroth order approximation to be perturbed. In our view, it is necessary to understand the principles behind individual effects before one can understand the interplay of effects\textemdash and what is the complex case but an interplay of individually simple effects? 

\section*{Acknowledgements}
We wish to thank Hart Goldman, Subir Sachdev and Piers Coleman for helpful and intriguing comments. 

\newpage
\appendix
\section{Free Fermion Green's Function in Bosonized Language}\label{appendix1}
As previous calculation, the calculation of the free fermion Green's function in the bosonized language requires that $\chi$ and the tangential position $x_{\parallel}$ be discretized. The free discretized action of a fermion is given by 
\begin{align}
     S_0[\zeta]=\dfrac{1}{8 \pi} \int dt\text{ } dx_{\perp} \sum_{n,m}  \dfrac{1}{v_F(\chi_n)}  &\left[\partial_{a} \zeta\left(t,x_{\perp},x_{\parallel,m},\chi_n\right)  \partial^{a} \zeta\left(t,x_{\perp},x_{\parallel,m},\chi_n\right)\right],
\end{align}
where a factor of $\frac{k_F}{2\pi}$ has \textit{not} disappeared\textemdash rather,  recall that the way in which it arose in the double scaling limit is through its relation to the the integration measure $\delta x_{\parallel}\delta \chi=\frac{2\pi}{k_F}$, and if not for the continuum limit it would not have appeared in the first place. In this case, the chiral electron-hole propagator is just a chiral Green's function of the $1+1$D wave equation, satisfying
\begin{equation}
     \dfrac{1}{2\pi}\partial_{\perp}\left(\partial_{t}-v_F\left(\chi_n\right)\partial_{\perp}\right) \Delta^{(0)}_{n,n,m,m^\prime}\left(t,t^\prime,x_{\perp},x^{\prime}_{\perp}\right)=\delta\left(t-t^\prime\right)\left(x_{\perp}-x_{\perp}^\prime\right) \delta_{m,m^\prime},
\end{equation}
whose general solution is defined up to a shift by a function of $x_{\parallel}-x^{\prime}_{\parallel}$, of which the equation is independent 
\begin{equation}
    \Delta^{(0)}_{n,n,m,0}\left(t,0,x_{\perp},0\right)=\log\left[x_{\perp}-v_F t+i\epsilon \text{sign}\left(x_{\perp}\right)\right]+C\left(x_{\parallel,m}\right).
\end{equation}
This shift ambiguity can be traced back to the chiral symmetry of the classical theory, under which $\zeta$ is shifted by an arbitrary function of $x_{\parallel}$. This results in a free electron Green's function defined up to a \textit{factor} of $x_{\parallel}$
\begin{align}
    \nonumber G^{(0)}_{+}\left(\boldsymbol{x},t\right)&=\exp\left(\Delta^{(0)}_{+}\left(\boldsymbol{x},t\right)\right)\\
    &=f\left(x_{\parallel}\right) \dfrac{1}{x_{\perp}-v_F t+i\epsilon \text{sign}\left(x_{\perp}\right)}.
\end{align}
The correct continuum limit is clearly obtained by setting  
\begin{align}
    \nonumber G^{(0)}_{+}\left(\boldsymbol{x},t\right)
    &=\delta\left(x_{\parallel}\right) \dfrac{1}{x_{\perp}-v_F t+i\epsilon \text{sign}\left(x_{\perp}\right)},
\end{align}
and, indeed, it clear from the action that variables at different $x_{\parallel}$ are not correlated. 

\section{Behavior of the Fourier Transform of the Long Comoving Distance Green's Function}\label{appendix2}
Let $u\equiv x_{\perp}-v_F t$. The Fourier transform of the long comoving distance Green's function, \eqref{elect-prop}, is given by

\begin{align}
    G_{+}(\omega,\boldsymbol{k})=\int^{\infty}_{-\infty} du \int^{\infty}_{-\infty} dx_{\perp} \dfrac{e^{i \left(\frac{\omega}{v_F}- k_{\perp}\right) x_{\perp}-i\frac{\omega}{v_F} u}}{u} \exp\left(-\gamma e^{-i\pi/6} \dfrac{\left|x_{\perp}\right|} {\left|u\right|^{2/3}}\right) 
\end{align}
The $x_{\perp}$ integral can be easily performed, giving
\begin{equation}\label{fourier}
    G_{+}(\omega,\boldsymbol{k})=\int^{\infty}_{-\infty} du \text{ }  {e^{-i\frac{\omega}{v_F} u}} \dfrac{2\gamma e^{-i\pi/6}}{u^{1/3}\left[\left(\omega-v_F k_{\perp}\right)^2 u^{4/3}+\gamma^2 e^{-i\pi/3} \right]}.
\end{equation}
One can then expand \eqref{fourier} in powers of $\omega$, but for constant $\omega-v_F k_{\perp}$ (it is natural to treat $\omega$ and $\omega-v_F k_{\perp}$, rather than $\omega$ and $k$, as independent variables since they do not intermix under the scaling transformation \eqref{scaling-momentum}). Clearly, the zeroth order term vanishes by antisymmetry in $u$. The first order term can be easily shown to diverge, consistent with an $\omega^{\alpha}$ behavior with some exponent $1>\alpha>0$, which by the scaling symmetry \eqref{scaling-momentum} should be given by $\frac{2}{3}$. Indeed, this can be confirmed by an explicit calculation. The small-$\omega$ propagator can be found by taking the limit of large $u$ in the integrand:
\begin{align}\label{fourier2}
    G_{+}(\omega,\boldsymbol{k})&\approxeq \int^{\infty}_{-\infty} du \text{ }  {e^{-i\frac{\omega}{v_F} u}} \dfrac{2\gamma e^{-i\pi/6}}{\left(\omega-v_F k_{\perp}\right)^2 u^{5/3}}\nonumber\\
    &=-B\dfrac{\text{sign}(\omega) \omega^{2/3}}{\left(\omega-v_F k_{\perp}\right)^2},
\end{align}
where 
\begin{equation}
    B=\dfrac{4}{\sqrt{3}} e^{-i\pi/3} \left(\dfrac{\pi^{4} g^{4}}{k_F v_F^4}\right).
\end{equation}

\section{Subdominance of Corrections Due to Quadratic Dipsersion}\label{appendix3}
For simplicity, we focus on $x_\parallel=0$, as is done in \cite{chubukov}, though the basic argument can be generalized to $x_{\parallel}\neq 0$. The long comoving-distance fermion propagator can then be written as
\begin{align}
     G_{+}(x_{\perp},x_{\parallel}=0,t)\propto\dfrac{1}{x_{\perp}-v_F t} \exp\left[-Q\left(t, x_{\perp}\right)\right],
\end{align}
for some function $Q$. In the linearized theory, we have obtained 
\begin{equation}
    Q\left(t, x_{\perp}\right)=-\gamma e^{-i\pi/6} \dfrac{\left|x_{\perp}\right|} {\left|x_{\perp}-v_F t\right|^{2/3}}.
\end{equation}
Chubukov et al.\cite{chubukov} suggest heuristically accounting for the effects of quadratic dispersion simply by adding it to the free electron propagators in \eqref{inter-prop}, i.e., by setting

\begin{equation}
    Q(\omega,k_{\perp})=
      - \int^{\infty}_{-\infty} \dfrac{dk_{\parallel}}{2\pi} \dfrac{4\pi^2 g^2}{\left(\omega-v_F k_{\perp}-\beta k_{\parallel}^2 \right)^2} \dfrac{1}{k^2_{\parallel}+i\dfrac{g^2 k_F}{4 \pi v_F}\dfrac{\left|\omega\right|}{\left|k_{\parallel}\right|}},
\end{equation}
with $\beta$ some coefficient which will be left unspecified so as to study the effects of neglecting it (physically, it scales as $k_F^{-1}$). The Fourier transform of $Q$ with respect to $k_{\perp}$ is the same as in \eqref{class-m} except with the frequency shifted by $-\beta k_{\parallel}^2$, giving
\begin{align}\label{curv-corr}
    Q(\omega,x_{\perp})&=-\dfrac{2\pi^2 g^2}{v_F ^2} \int^{\infty}_{-\infty} \dfrac{dk_{\parallel}}{2\pi}  \dfrac{1}{k^2_{\parallel}+i\dfrac{g^2 k_F}{4 \pi v_F}\dfrac{\left|\omega\right|}{\left|k_{\parallel}\right|}}  \left|x_{\perp}\right| \exp\left[\dfrac{i}{v_F}\left(\omega-\beta k_{\parallel}^2\right) x_{\perp}\right] \nonumber\\
    &=-\dfrac{2\pi^2 g^2}{v_F ^2}\left(\dfrac{g^2 k_F}{4\pi v_F} \omega\right)^{-1/3} \left|x_{\perp}\right| F\left(i\dfrac{\beta}{v_F}  \left(\frac{g^2 k_F}{4\pi v_F } \omega\right)^{2/3}x_{\perp}  \right) \exp\left(\dfrac{i}{v_F}\omega x_{\perp}\right),
\end{align}
where
\begin{equation}
    F\left(z\right)\equiv \int^{\infty}_{-\infty}\dfrac{ds}{2\pi} \dfrac{1}{s^2+\frac{i}{\left|s\right|} }\exp\left(-s^2 z\right).
\end{equation}

Since $F$ depends on $\omega^{2/3}$ and $\beta$ solely through their product, it is immediately clear that the small frequency limit is equivalent to the small $\beta$ limit\footnote{It might seem inconsistent to neglect $\omega^{2/3}$ in the argument of $F$ while keeping the factor of $\omega x_{\perp}$ in the exponent. However, the sole effect of the latter is to replace time with comoving distance. It is certainly ``legal'' to neglect this latter factor, which amounts to replacing $\omega-v_F k_{\perp}$ with $-v_F k_{\perp}$ and will have no effect on the $\omega^{2/3}$ self-energy.}:
\begin{equation}
    Q(\omega,x_{\perp})\approxeq -\dfrac{2\pi^2 g^2}{v_F ^2}\left(\dfrac{g^2 k_F}{4\pi v_F} \omega\right)^{-1/3} \left|x_{\perp}\right| F\left(0\right),
\end{equation}
i.e., the low-energy contribution to $Q$ is exactly that obtained within the linearized theory. 

The low-energy contributions to $Q$ will govern the long-time behavior of the propagator, and thus the low-energy behavior \textit{of the propagator}. It thus follows that the low-energy propagator is unaffected by the quadratic dispersion. This can be confirmed by computing $F(z)$ exactly\footnote{Note that it is necessary to compute the integral first and only then expand in $\omega$ rather than the other way around. The reason is that since the expansion of $F(z)$ involves fractional powers of $z$, its derivatives diverge at $z=0$.}, which can be done on a computer. The result is 
\begin{align}
    \nonumber F(z)=&-\dfrac{1}{\sqrt{\pi}}\prescript{}{1}F_3\left(1;\dfrac{1}{2},\dfrac{5}{6},\dfrac{7}{6};-\dfrac{1}{27} z^3\right)\\&+\dfrac{1}{6} e^{-z/2}\left[e^{3z/2}+\cos\left(\dfrac{\sqrt{3} z}{2}\right)+\sin\left(\dfrac{\sqrt{3} z}{2}\right)\right],
\end{align}
where $\prescript{}{p}F_q$ is (again) the generalized hypergeometric function. $Q\left(\omega,x_{\perp}\right)$ can then be expanded for small frequencies, and since each term is just a function $\omega$ it can be easily transformed to the time domain. The leading order correction due to the quadratic dispersion is of the form
\begin{equation}
    Q^{(1)}\left(t,x_{\perp}\right)\propto \beta^{1/2} \text{sign}\left(x_{\perp}\right) \delta\left(x_{\perp}-v_F t\right) \left(x_{\perp}\right)^{3/2},
\end{equation}
which is just a delta function in the comoving distance. For $n>1$, the $n$th order correction is of the form
\begin{equation}
    Q^{(n)}\left(t,x_{\perp}\right)\propto \beta^{n/2} \text{sign}\left(x_{\perp}\right) \left(x_{\perp}-v_F t\right)^{-\frac{1}{3}(2+n)} \left(x_{\perp}\right)^{1+n/2}.
\end{equation}
It is thus clear that all the corrections due to quadratic dispersion decay faster at long times than the result of the linearized theory. They will thus correspond to corrections to the self energy of higher orders in $\frac{\omega^{2/3}}{\omega-v_F k_{\perp}}$.

\newpage
\printbibliography

\end{document}